\documentclass[]{JHEP3} 

\let\ifpdf\relax
\usepackage{ifpdf}

\usepackage{epsfig,multicol,bbm}
\usepackage[sort&compress,numbers]{natbib}

\newcommand\fverb{\setbox\fverbbox=\hbox\bgroup\verb}
\newcommand\fverbdo{\egroup\medskip\noindent%
			\fbox{\unhbox\fverbbox}\ }
\newcommand\fverbit{\egroup\item[\fbox{\unhbox\fverbbox}]}
\newbox\fverbbox

\usepackage{graphicx}

\newcommand{\beq}{\begin{equation}}
\newcommand{\eeq}{\end{equation}}
\newcommand{\beqa}{\begin{eqnarray}}
\newcommand{\eeqa}{\end{eqnarray}}

\def\nn{\\ \nonumber}

\def\cal{\mathcal}

\def\met{\ifmmode%
\setbox0=\hbox{$E_t$}%
\setbox1=\hbox to\wd0{\hss$/$\hss}\else%
\setbox0=\hbox{E_t}%
\setbox1=\hbox to\wd0{\hss/\hss}\fi%
E_t\hskip-\wd0\box1 }

\title{The Matrix Element Method at Next-to-Leading Order.}

\author{
    John M. Campbell, Walter T. Giele and Ciaran Williams
    \\
    Fermilab, Batavia, IL 60510, USA
    \\
    E-mails: 
    {\tt johnmc@fnal.gov}, 
    {\tt giele@fnal.gov}, 
    {\tt ciaran@fnal.gov}.}

\received{\today} 		
\accepted{\today}		

\preprint{
FERMILAB-PUB-12-087-T}

\abstract{This paper presents an extension of the matrix
element method to next-to-leading order 
in perturbation theory, 
for electro-weak final states. To accomplish this we have developed a method to calculate 
next-to-leading order weights on an event-by-event basis. 
This allows for the definition of next-to-leading order
likelihoods in exactly the same fashion as at leading order, thus extending
the matrix element method to next-to-leading order.
A welcome by-product of the method is the straightforward and efficient
generation of unweighted next-to-leading order events.
As examples of the application of our next-to-leading order matrix element method we consider the
measurement of the mass of the $Z$ boson and also the search for the Higgs boson in the four lepton channel.   
}


\begin{document}


\section{Introduction} 

The continued successful running of the LHC is already resulting in an impressive 
data set with which to test the Standard Model (SM). One of the main 
aims of the experimental program is to observe the mechanism behind electroweak 
symmetry breaking, for which the postulated Higgs boson is a theoretically well-motivated 
example. Using the 5 fb$^{-1}$ data set the LHC has tightly constrained the mass of the Higgs boson,
whilst also providing tantalising hints in the low mass region ($\sim 120-125$~GeV)~\cite{ATLAS:2012ae,Chatrchyan:2012tx}. 
Present analyses often use data driven techniques for background estimation with an emphasis on accurate signal modeling,
for instance in the diphoton Higgs search~\cite{ATLAS:2012ad,Chatrchyan:2012tw}. Whilst this is 
a sensible strategy for searches, after discovery an accurate modeling of both signal and background will be required 
in order to confirm the exact properties of any new particle, such as its spin and couplings~\cite{Gao:2010qx,DeRujula:2010ys}. 
In addition to Higgs searches, precision measurements in the electroweak sector of the SM could also provide valuable insight.
By measuring top quark properties and electroweak gauge boson couplings, potential new physics contributions can be constrained.
A recent example, that exhibits some tension with the SM, is provided by measurements of the top quark forward-backward
asymmetry at the Tevatron~\cite{Aaltonen:2011kc,Abazov:2011rq}.

There are many methods available for performing studies of particle properties, for instance for measuring
their masses or investigating their interactions. Among these, the matrix element method (MEM) stands out
since it is sensitive to all the available kinematic information for each individual event.
Originally pioneered at the Tevatron~\cite{Kondo:1991dw,Dalitz:1991wa}, the MEM has proven extremely useful in the
the top sector~\cite{Abazov:2004cs,Abazov:2006bd,Abulencia:2007br,Abulencia:2006ry,Abazov:2006bg,Fiedler:2010sg}. Recently the method has been
used to observe single top production~\cite{Abazov:2008kt,Aaltonen:2010jr,Aaltonen:2009jj,Abazov:2009ii} and to provide evidence for
top quark spin correlations~\cite{Abazov:2011gi}. The MEM has also been used to try to improve searches for the Higgs boson
in the associated production channel~\cite{Aaltonen:2011rt}.
At the LHC the MEM is also beginning to be used, for example in the measurement 
of the electroweak mixing angle at CMS~\cite{Chatrchyan:2011ya}.

The popularity of the MEM is based on its ability to utilize the theoretical prediction from the matrix element, retaining
all the hard scattering correlations.
For each experimental event, the MEM assigns a probability that it can be described by a given theoretical model.
In this way one can produce a likelihood that the theoretical model
describes a particular set of data. 
Matrix elements (at tree-level) are relatively straightforward to calculate and automated tools for this purpose 
have been available for several years~\cite{Alwall:2011uj,Gleisberg:2008fv,Kilian:2007gr,Boos:2004kh,Mangano:2002ea}. 
Indeed, the application of automated tools to the MEM was previously considered in ref.~\cite{Artoisenet:2010cn}.
However, a serious limitation of the method is that it has so far been
defined only at leading order (LO). For the precision studies that will become possible with the wealth of
data at the LHC, it is crucial to extend and adapt the method such that it is defined at higher orders.
An implementation of the method at next-to-leading order (NLO), the de facto standard for most theoretical predictions at the LHC,
is required to put the MEM on a solid theoretical footing and elevate the method to being a robust analysis tool.

The absence of higher order corrections in current implementations of the MEM is easily understood.
It is not immediately clear how to use existing NLO calculations to associate a NLO weight with a given exclusive experimental event.
This is primarily due to the fact that NLO calculations include contributions from
both loop and bremsstrahlung diagrams, that must be integrated over different physical phase spaces.
As such there is no clear one-to-one map between an exclusive event, containing a finite number of objects with measured properties,
and a NLO weight.  Addressing this very issue is the principal goal of this paper.

We therefore present a method of calculating NLO weights suitable for use with the MEM approach. 
As a welcome by-product, the method also provides a procedure for calculating unweighted NLO events. 
As a first step, in this paper we consider only the production of 
colour neutral final states. This ensures that at NLO the real phase space is associated with radiation from initial state partons only.
We thus postpone the treatment of final state jets at NLO to a future publication.

This paper proceeds as follows. In section~\ref{sec:MemLO} we first introduce the MEM at LO and discuss its use in experimental 
analyses. Section~\ref{sec:MemNLO} explains our extension of the MEM
to NLO 
and discusses the generation of unweighted NLO events.
In section~\ref{sec:valDY} we validate the 
code using MCFM~\cite{Campbell:1999ah,Campbell:2010ff,Campbell:2011bn,MCFMweb} and Pythia~\cite{Sjostrand:2007gs}. 
Section~\ref{sec:ZZ} is devoted to an application of immediate phenomenological
interest, namely the search for a Higgs boson in the $ZZ^\star$ decay channel to
four leptons. Finally in 
section~\ref{sec:conclusions} we draw our conclusions. The appendices describe the generation of the 
phase space in more detail and discuss the modifications to the usual dipole subtraction procedure that are required in
our approach.

\section{The Matrix Element Method at Leading Order}
\label{sec:MemLO}

In this section we define the MEM at LO and discuss how it may be
used in experimental analyses.

\subsection{Overview of the MEM}

We begin by assuming that one wishes to measure a model parameter $\Omega$, using an experimental data
set $\{{\bf{x}}\}$ that contains $N$ events ${\bf{x_i}}$.  One method to determine the best-fit value of
$\Omega$ is to construct 
a probability density function in which each event is weighted by the LO scattering
probability computed with the parameter $\Omega$. The resulting probability density
function associated with a single event $\bf{x}$, for a given $\Omega$, can be written schematically as,
\begin{eqnarray}
\mathcal{P}({\bf{x}}|\Omega)=\frac{1}{\sigma^{LO}_{\Omega}}\int
dx_adx_b\,d{\bf{y}} \sum_{ij}
\frac{f_i(x_a)f_j(x_b)}{x_ax_b s} 
\, \mathcal{B}^{ij}_{\Omega}(p_a,p_b,{\bf{y}}) \, W({\bf{x}},{\bf{y}})\ .
\label{eq:MEM}
\end{eqnarray}
In this equation $f_i(x_a)$ and $f_j(x_b)$ represent the parton distribution functions for
partons of flavours $i$ and $j$ possessing momentum fractions $x_a$ and $x_b$ of their parent hadrons.
$\mathcal{B}^{ij}_{\Omega}(p_a,p_b,{\bf{y}})$ is the LO scattering probability with partons $i$ and $j$ in the initial state. 
The hadron collision takes place at a centre of mass energy  $\sqrt{s}$ while the flux factor
entering in the denominator of Eq.~(\ref{eq:MEM}) is the partonic centre of mass
energy squared, $s_{ab} = x_a x_b s$.

An experimental event ${\bf{x}}$ is by definition a detector level event, whilst the scattering probability 
is computed theoretically at the level of partons. Therefore in order to correctly use the scattering probability
as a probability density function  one must include effects that model this discrepancy. 
The transfer function $W({\bf{x}},{\bf{y}})$ relates a detector level
event ${\bf{x}}$ to a particle level event ${\bf{y}}$ that can be used to compute the scattering amplitude.
This transfer function, dependent on the specifics of the experimental set-up,
takes account of factors such as limitations on the energy resolution and acceptance of the detector. 
The transfer function is constructed such that it is itself a probability density function,
\beq
\int d{\bf{y}} \, W({\bf{x}},{\bf{y}})=1 \;.
\label{eq:transfernorm}
\eeq 
Finally, the factor $\sigma_{\Omega}$ is  the total cross section for the process for a specific
choice of $\Omega$, thus ensuring that the probability distribution is properly normalized to unity. 

Once the probability density function $\mathcal{P}({\bf{x}}|\Omega)$ has been computed for each
event $\bf x$, it is straightforward to compute a likelihood for the data set as a whole.
For the data set $\{\bf{x}\}$ with $N$ events, the likelihood function 
$\mathcal{L}(\{{\bf{x}}\}|\Omega)$ for a given parameter $\Omega$ is defined by, 
\begin{eqnarray}
\mathcal{L}(\{{\bf{x}}\}|\Omega) = f(N) \prod_{i=1}^N \mathcal{P}({\bf{x_i}}|\Omega).
\label{eq:Likdefn}
\end{eqnarray} 
Here $f(N)$ is a normalisation factor related to the overall number of events in the data set. 
In most analyses one is interested in comparing two hypotheses, either in 
the form of a likelihood ratio, or more commonly by comparing the difference of two log-likelihoods. 
Therefore in most practical applications the explicit form of $f(N)$ is unimportant. This is the case
for all the examples that we present here and, as such, we will simply drop the factor $f(N)$ in 
Eq.~(\ref{eq:Likdefn}).

By construction, the value of the likelihood function will be larger for theories that describe the data better. 
The best fit corresponds to the parameter choice $\Omega$ that maximises $\mathcal{L}$ (and hence also
$\log{\mathcal{L}}$). In the region of the maximum -- and as long as the data set is large enough --
departures from the maximum value of the likelihood can be simply interpreted in terms of standard
deviations from the best fit. Since we consider a single parameter $\Omega$, the likelihood can be described by a parabola
in the region of the maximum (see e.g. ref~\cite{Cowan:stat}) and standard deviations (here represented by $n\sigma$) from the observed maximum 
can then be defined by,
\begin{eqnarray}
 \log{\mathcal{L}} \left.\right|_{n\sigma} =\log{\mathcal{L}_{max}}-n^2/2. 
\label{eq:nsig}
\end{eqnarray}
In our examples we will use this to define one- and two-sigma confidence levels for our results, although we stress that our
studies do not include detector effects and are thus only for the sake of illustration.

\subsection{Leading order formulation}

We now return to the probability density function, Eq.~(\ref{eq:MEM}). We recall that at this order,
\beq
\mathcal{B}^{ij}_{\Omega}(p_a,p_b,{\bf{y}})=|\mathcal{M}^{ij,(0)}_{\Omega}(p_a,p_b,{\bf{y}})|^2\;,
\eeq
where $\mathcal{M}^{ij,(0)}_{\Omega}({\bf{y}})$ is the leading order matrix element for the relevant process with initial state partons $i$ and $j$.
In this paper we make the simplifying assumption that the events completely specify the final state particles so
that, for example, we do not consider events containing neutrinos. 
For a Born point $p$ the constraint of momentum conservation fixes the values of the parton fractions
$x_a$ and $x_b$. By convention we position the incoming particles along the $z$-axis in the lab frame and then use the
momentum conserving delta function between the $n$ final-state particles $\{p_1,\ldots,p_n\}$,
\begin{eqnarray}
\delta^{(4)}(p_a+p_b-\sum_{i=1}^{n}p_i) \;,
\end{eqnarray}
to find the relations,
\beq
x_a-x_b=\frac{2}{\sqrt{s}} \left(\sum_{i=1}^{n}p_i^z\right) \;, \qquad 
x_a+x_b=\frac{2}{\sqrt{s}} \left(\sum_{i=1}^n E_i\right) \;.
\label{eq:x1x2def}
\eeq 

\begin{figure} 
\begin{center}
\includegraphics[width=14cm]{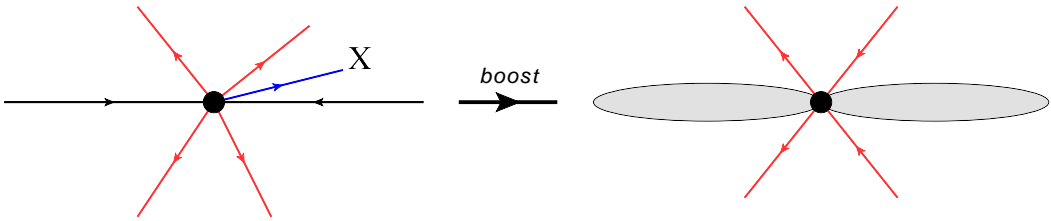}
\caption{The generation of the Born (and virtual) phase space from a given experimental event. The left hand side depicts a collision that results in the production 
of a colour neutral final state (represented here by four leptons in red) that do not balance in the transverse plane. The resulting imbalance ($X$, in blue) represents
the remaining event which is not modelled in the Born matrix element. We apply a Lorentz transformation such that $X$ has no components in the transverse plane,
with the remaining longitudinal and energy components absorbed into the colliding partons.} 
\label{fig:BornBoost}
\end{center}
\end{figure}

However, matching an experimental point $\tilde{p}$ to the LO kinematics $(p)$ is a challenge. In particular,
any event will always contain  additional radiation that is not modelled by the leading order (Born level) matrix element.
In order to proceed we shall define a four vector $X$, that balances the momenta of the final state particles.
This is illustrated schematically in Fig.~\ref{fig:BornBoost} and expressed through the equations,
\begin{eqnarray}
X = -\sum_{i=1}^{n} \tilde{p}_{i}. 
\end{eqnarray} 
The Born matrix elements, with the beam directions consistently along the $z$-axis, are only defined for $X^x=X^y=0$,
i.e. when there is no $p_T$ imbalance between the final state particles\footnote{Attempting 
to evaluate a LO matrix element with a phase space point that does not conserve momentum is ill-defined. The exact weight obtained depends on which 
kinematic invariants one has chosen to use in the expression for the matrix element.}.
Therefore, in order to ensure that the experimental event has a well-defined interpretation as a Born level
phase space point we need to remove the transverse components of $X$. This can be achieved by applying a Lorentz transformation
$\Lambda(X)$ on the momenta $\tilde{p}$ in the event to arrive at a frame in which the transverse components of $X$ are zero, 
\begin{eqnarray}
p_i^\mu=\Lambda^\mu_{~\nu}(X) \, \tilde{p}_i^\nu \qquad {\rm{with}}
 \qquad \sum_{i=1}^n p_i^x = \sum_{i=1}^n p_i^y = 0  \;.
\label{eq:boostdef}
\end{eqnarray}
As desired, the phase space point $p$ is now of the correct form to be used in a Born level matrix element. For a given transformation,
the momentum fractions $x_a$ and $x_b$ are then related to the transformed momenta $p $ through
the relations in Eq.~(\ref{eq:x1x2def}). However, we note that Eq.~(\ref{eq:boostdef}) does not specify a unique transformation.
We can define multiple transformations that result in $X^x=X^y=0$ and
that yield different longitudinal components of $p$. In other words $x_a$ and $x_b$ are frame-dependent
quantities determined by the boost choice and it is only the product $x_a x_b$ that is Lorentz invariant.
Therefore in order to produce a sensibly defined 
weight for each event we must integrate over this unobservable degree of freedom. 

To illustrate these ideas in more detail we begin with the usual definition of the total cross section for the production of $n$ massless final state particles,  
\begin{eqnarray}
\sigma^{LO}_{\Omega}=(2\pi)^{4-3n}\int dx_a\,dx_b\prod_{m=1}^{n} \bigg(\frac{d^3 {\bf{p}}_{m}}{2E_m}\bigg)  \frac{f_i(x_a)f_j(x_b)}{x_ax_b s}\,\mathcal{B}^{ij}_{\Omega}\,\delta^{(4)}\left(p_a+p_b-\sum_{i=1}^{n} p_i\right).\label{eq:LO_XS}
\end{eqnarray}
Here we have suppressed the dependence of $\mathcal{B}$ on the kinematics and the summation over $i$ and $j$ for clarity. We wish to factorise Eq.~(\ref{eq:LO_XS}) into 
two pieces, one representing initial state production and the other the decay of a heavy object into the final state particles. To this end we define $Q=p_a+p_b$
and insert the operator $\int dQ^2\, \delta(x_ax_bs-Q^2)=1$, 
\begin{eqnarray}
\sigma^{LO}_{\Omega}&=&(2\pi)^{4-3n}\int dx_a\,dx_b\,dQ^2\, \delta(x_ax_bs-Q^2) \nonumber \\
&\times& \prod_{m=1}^{n} \bigg(\frac{d^3 {\bf{p}}_{m}}{2E_m}\bigg)
 \frac{f_i(x_a)f_j(x_b)}{x_ax_b s}\mathcal{B}^{ij}_{\Omega} \,\delta^{(4)}\left(Q-\sum_{m=1}^np_m\right) \;.
\label{eq:LO_XS_II}
\end{eqnarray}
For the remainder of this paper we will define the phase space element associated with the final state particles as, 
\begin{eqnarray}
d{\bf{x}}= (2\pi)^{4-3n}dQ^2 \prod_{m=1}^{n} \bigg(\frac{d^3 {\bf{p}}_{m}}{2E_m}\bigg)\delta^{(4)}\left(Q-\sum_{m=1}^n p_m\right) \;.
\end{eqnarray} 
Using this definition we see that,
\begin{eqnarray}
\sigma^{LO}_{\Omega}&=&\int dx_a\,dx_b\,d{\bf{x}}\, \delta(x_ax_bs-Q^2)\frac{f_i(x_a)f_j(x_b)}{x_ax_b s}\mathcal{B}^{ij}_{\Omega}(p_a,p_b,\bf x) \;. \nonumber\\
&=&\int \,d{\bf{x}}\,  \mathcal{L}_{ij}(Q^2,x_l,x_u)
\mathcal{B}^{ij}_{\Omega}(p_a,p_b,{\bf {x}}).
\label{eq:sigmaLO}
\end{eqnarray}
This separation is convenient since $\mathcal{B}^{ij}_{\Omega}(p_a,p_b,\bf x)$ is Lorentz invariant and need only 
be evaluated for a single phase space point. 
The process independent integration over boosts is given by, 
\begin{eqnarray}
\mathcal{L}_{ij}(s_{ab},x_l,x_u) &=& \int dx_a dx_b \, \frac{f_i(x_a)f_j(x_b)}{x_ax_bs} \, \delta(x_ax_bs-s_{ab}) \nonumber \\
&=&  \int_{x_l}^{x_u} dx_a \, \frac{f_i(x_a)f_j(s_{ab}/(sx_a))}{sx_a s_{ab}} \;,
\label{eq:tauint}
\end{eqnarray}
where in the second expression we have made the dependence on the upper and lower bounds explicit. 

This factorisation in terms of initial and final state variables is exactly what we require to build our probability density function for the MEM since the experimental 
input is always a final state phase space point ${\bf{x}}$. We  can define Eq.~(\ref{eq:MEM}) more formally as,  
\beq
\mathcal{P}({\bf{x}}|\Omega)=\frac{1}{\sigma^{LO}_{\Omega}}\int \,d{\bf{y}}\,\mathcal{L}_{ij}(s_{ab},x_l,x_u)
\mathcal{B}^{ij}_{\Omega}(p_a,p_b,{\bf {y}})W({\bf{x}},{\bf{y}}) \;.
\label{eq:boostMEM}
\eeq

For a completely inclusive description of the final state, Eqs.~(\ref{eq:tauint}) and~(\ref{eq:boostMEM}) are sufficient. However, realistic
applications require transverse momentum and pseudo-rapidity cuts in order to define fiducial regions of the detector.
It is therefore useful to consider the forms of the lab frame transverse momentum ($p_T^{{\tiny lab}}$) and pseudo-rapidity ($\eta^{{\tiny lab}}$)
under the application of a given longitudinal boost parameterized by $x_a$.

The four-momenta of all the particles depend on the boost parameter -- the initial state momenta $p_a(x_a)$, $p_b(x_a)$ and
the momentum of particle $i$ in the final state, $p_i(x_a)$. However we note that invariant masses,
$s_{ij} = 2 p_i(x_a) \cdot p_j(x_a)$ cannot depend on the boost and may therefore be evaluated using any choice of boost
parameter. The lab frame transverse momentum and pseudo-rapidity are defined in
terms of such invariants and the boost parameter $x_a$ by,
\beqa\label{invariants}
p_T^{{\tiny lab},i}&=&\sqrt{\frac{s_{ai}s_{ib}}{s_{ab}}} \;, \quad
\eta^{{\tiny lab},i}=\frac{1}{2}\log\left(\frac{x_a^2s}{s_{ab}}\frac{s_{ib}}{s_{ai}} \right).
\label{eq:ptydef}
\eeqa 
From these equations we see that $p_T^{{\tiny lab},i}$ does not depend on the boost parameter and therefore cuts on this
quantity can be performed outside the boost integration, i.e. in Eq.~(\ref{eq:boostMEM}). On the other hand, $\eta^{{\tiny lab},i}$
depends on $x_a$, so that cuts on the lab frame pseudo-rapidity should be included in Eq.~(\ref{eq:tauint}). These cuts constrain the
range of allowed boosts, i.e. the integration limits $x_l$ and $x_u$ are fixed by $|\eta_{max}|$.

In summary, by boosting an event to a frame in which the final state is $p_T$-balanced we have recovered Born kinematics and can
assign a likelihood to the event uniquely. 
Frequently in the next sections we will refer to these frames, in which the Born event is well defined, as the ``MEM frame''. As we have
discussed, this definition is only unique in the transverse plane and the ``MEM frame" is actually a set of equivalent frames connected
by longitudinal boosts. 

For the remainder of the paper we will make a simplification by assuming a ``perfect'' detector, i.e. the
transfer function is equal to  $W(\bf{x},\bf{y})=\delta({{\bf{x}}-{\bf{y}}})$.
This assumption is only valid for well-measured final state particles such as leptons and 
therefore as examples we only consider $ZZ\rightarrow 4\ell$ and $Z \rightarrow \ell^+\ell^-$. 
One may worry that the additional integrations imposed by the transfer functions spoil the method. 
In particular the transfer functions are defined both in a specific frame and given detector setup. However 
the construction of the MEM allows for a convenient factorisation of the problem. The role of the transfer functions is to 
provide a model describing the range of possible particle level events which could be generated given a specific detector 
event. Therefore even though the transfer functions are non-Lorentz invariant (as indeed are the PDFs) 
they do not spoil the method. They merely result in one event being replaced by an integral over many similar events, which 
in turn each get boosted to the MEM frame and analysed individually. The total weight for one experimental event is thus 
obtained by performing these additional integrations. The only remaining caveat is to correctly normalise the sample by including 
the transfer functions in the cross section definition. 
\begin{eqnarray}
\sigma^{LO,trans}_{\Omega}&=&\int {d\bf y}\, dx_a\,dx_b\,d{\bf{x}}\, \delta(x_ax_bs-Q^2)\frac{f_i(x_a)f_j(x_b)}{x_ax_b s}\mathcal{B}^{ij}_{\Omega}(p_a,p_b,{\bf x}) W({\bf{x}},{\bf{y}}) \;.
\end{eqnarray}
Taking this simplification and the integration over the longitudinal boost into account,
Eq.~(\ref{eq:boostMEM})  becomes,
\beq
\mathcal{P}({\bf{x}}|\Omega)=\frac{1}{\sigma^{LO}_{\Omega}}\mathcal{L}_{ij}(s_{ab},x_l,x_u)
\mathcal{B}^{ij}_{\Omega}(p_a,p_b,{\bf{x}})\ .\label{eq:MEM_dirx}
\eeq
The above equation defines the LO probability density function for the MEM. We recall that $\mathcal{B}^{ij}_{\Omega}(p_a,p_b,{\bf{x}})$ represents 
the Born Matrix element squared, $|\mathcal{M}^{ij,(0)}_{\Omega}(p_a,p_b,{\bf{x}})|^2$ and that $\sigma_{\Omega}$ represents the fiducial cross section,
calculated using cuts in the lab frame. We define the following quantity, 
\begin{eqnarray}
B_{\Omega}({\bf{x}})=\mathcal{L}_{ij}(s_{ab},x_l,x_u)\,\mathcal{B}^{ij}_{\Omega}(p_a,p_b,{\bf{x}}) \;,
\end{eqnarray} 
and observe from Eq.~(\ref{eq:sigmaLO}) that $\int d{\bf{x}} \,B_{\Omega}({\bf{x}}) = \sigma^{LO}_{\Omega}$.  We can thus simplify Eq.~(\ref{eq:MEM_dirx}) to, 
\begin{eqnarray} 
\mathcal{P}({\bf{x}}|\Omega)=\frac{1}{\sigma^{LO}_{\Omega}} \, B_{\Omega} ({\bf{x}}).
\label{eq:P_LO} 
\end{eqnarray}
This formalism will prove useful in the following section when we extend the MEM to NLO. 

Using the techniques outlined above we have defined a procedure that takes an observed final state,
$\tilde{Q} + X$ and relates it to a LO model for the process, $p_a+p_b \rightarrow Q$. Specifically,
given an arbitrary amount of additional radiation we create a phase space point that recovers the Born kinematics, 
at the cost of introducing an integration over the longitudinal degree of freedom. 

Clearly this model will be better for events in which the momentum imbalance $X$ is small, rather than events 
in which $X$ is kinematically relevant, i.e. in the presence of one or more additional jets. When additional jets are present 
one has three options. The first option is to simply apply the LO model presented above, boosting the 
jet into the initial state. Since in general one expects this method to be rather sensitive to the amount of radiation, i.e. the
transverse momentum of the jet, it is prudent to check the validity of this approach by also considering smaller data sets obtained 
by applying a jet veto. If there are sufficient events, restricting the data set by imposing a strict jet veto is preferred since,
by ensuring that no additional hard jets are present, one can be confident that the LO model works reasonably well.
We shall present an example of applying such a jet veto in section~\ref{sec:valDY}. 
The second option is to use a LO calculation that already contains an additional jet, i.e. $p_a+p_b \rightarrow 
Q+\mbox{jet}+X$. In this case the extra radiation is well modelled but the MEM must be extended to include a systematic treatment of jets. In this paper 
we will not consider this option further.

Finally, one may try to systematically improve the MEM in an attempt to model the additional radiation. 
This is the approach discussed in Ref.~\cite{Alwall:2010cq}, with reference to initial state radiation. 
Instead one may incorporate such effects by extending the MEM to NLO. Since a NLO calculation includes the 
radiation of one additional parton, a first approximation of the effects of further radiation is made at this order. 
In the next section we will illustrate how this may be achieved within the MEM framework. 

Before extending the MEM to NLO we discuss potential differences between our implementation of the MEM and 
one in which there is no integration over the longitudinal degrees of freedom. One could imagine defining $x_1$ and $x_2$ 
from the input event and using these values in the PDF evaluation. Note however that this is only theoretically well defined 
in the limit of Born kinematics, i.e. it is only well-defined if the input event is LO. We have compared our implementation to one 
which does not integrate over the PDFs using LO events from MCFM. We find, as expected that in both cases the two cases 
give the same best fit value (in this case for $m_Z$) with the same errors. Note that such comparisons can only be made at LO, since 
the definition of $x_1$ and $x_2$ in the non-integration method is ambiguous.

\section{The Matrix Element Method at Next-to-Leading Order}
\label{sec:MemNLO}

In this section we define the MEM at NLO and, as a by-product, discuss how
one may generate unweighted events at NLO. 

\subsection{Going beyond LO: Defining NLO on an event by event basis}

The goal of this sub-section is to illustrate how to extend the MEM to NLO in perturbation theory.
However this is not a simple task since in a normal NLO calculation virtual and bremsstrahlung events live in separate phase spaces, their 
only communication being through a regularising subtraction scheme.
Instead of following this procedure, we need to reorganise the calculation such that it can provide a NLO weight
for a given Born event, with the sum over the event weights
recovering the usual NLO cross section.  To do this we begin by assuming that our event has been rendered in the MEM frame 
using the procedure described in the previous section. We note however that the procedure we will outline in this section is not useful solely 
for extending the MEM to NLO. We are creating a method for producing a NLO cross section from a series of Born phase space points, a
procedure that may have broader applications than are presented here.

Given the phase space point ${\bf{x}}=p_1,\dots, p_n$ where the final state momenta are those 
of the identified final state particles, we can define the NLO corrections by,
\begin{eqnarray}
\frac{d\,\sigma^{NLO}_{\Omega}({\bf{x}})}{d{\bf{x}}} = R_{\Omega}({\bf{x}}) + V_{\Omega}({\bf{x}}) \;.
\label{eq:Kfac}
\end{eqnarray}
This follows the usual separation of the NLO calculation into two pieces, each of which is associated with a different phase space. We stress though
that here the separation has been performed for a fixed Born phase space point, ${\bf x}$.
The definition of the term associated with the virtual corrections is straightforward since it is defined in the same phase space as the Born
contribution. Explicitly, we can define $V_{\Omega}({\bf{x}})$ as, 
\begin{eqnarray}
V_{\Omega}({\bf{x}})&=&\mathcal{L}_{ij}(s_{ab},x_l,x_u)\bigg(
\mathcal{B}^{ij}_{\Omega}(p_a,p_b,{\bf{x}})+\mathcal{V}^{ij}_{\Omega}(p_a,p_b,{\bf{x}})\bigg)\nonumber\\
&&+\sum_{m=0}^2 \int dz \bigg(\mathcal{D}_m(z,{\bf{x}}) \otimes\mathcal{L}_{m}(z,s_{ab},x_l,x_u)\bigg)_{ij}
 \mathcal{B}^{ij}_{\Omega}(p_a,p_b,{\bf{x}}).
\label{eq:kvirt}
\end{eqnarray} 
Here the first term represents the combination of the Born matrix element $\mathcal{B}^{ij}_{\Omega}$ and the one-loop Born interference 
term $\mathcal{V}_{\Omega}=2\mbox{Re}|\mathcal M^{(0)*}_{\Omega}M^{(1)}_{\Omega}|$
(where the dependence on the initial state partons has been suppressed).  This is coupled to the same boost function, $\mathcal{L}_{ij}$
as was defined at LO. 
In our approach we have followed the NLO implementation of MCFM and used the dipole subtraction procedure of Catani and Seymour~\cite{Catani:1996vz}
to handle the singularities in the virtual and real calculations. 
The final term in Eq.~(\ref{eq:kvirt}) contains the integrated subtraction terms, $\mathcal{D}_a$, introduced in this formalism.
Since we are considering initial state singularities the 
integrated dipoles depend on a convolution variable $z$. This variable is convoluted with the boost function to create three structures, 
\begin{eqnarray}
\mathcal{L}_0=\mathcal{L}, \quad \mathcal{L}_1=  \int_{x_l}^{x_u} dx_a \, \frac{f_i(x_a/z)f_j(s_{ab}/(sx_a))}{zs x_a s_{ab}}, \quad  \mathcal{L}_2=  \int_{x_l}^{x_u} dx_a \, \frac{f_i(x_a)f_j(s_{ab}/(zsx_a))}{zs x_a s_{ab}}. \nonumber\\
\end{eqnarray}
In Eq.~(\ref{eq:kvirt}) the sum over these convolutions is given by $m$.

Using Eq.~(\ref{eq:kvirt}) we are able to define an event by event finite weight associated with the Born plus virtual contributions. 
Our remaining task is thus to define $R_{\Omega}({\bf{x}})$ such that there is no double counting of events. In other 
words we must ensure that the integration of Eq.~(\ref{eq:Kfac}) results in the total NLO cross section $(\sigma^{NLO}_{\Omega})$. 
One way to ensure this is to use a forward branching phase space generator (FBPS) \cite{Giele:2011tm} to construct the real phase space. 
Starting from the Born phase space point, $\hat p_a+\hat p_b\rightarrow Q$ the FBPS generates the real radiation by branching one of the 
initial state momenta to produce the real phase space point $p_a+p_b\rightarrow Q+p_r$. 
In the following we will use the hatted notation to indicate 
a Born phase space point, whilst the un-hatted momenta represent the real phase space point. 

The phase space generator needs to integrate out all initial state radiation
within the constraints of fixed momenta of the identified final state particles (and, if required, the jet veto).
We show in Appendix~\ref{app:FBPS} that this can be achieved using a FBPS generator defined by,
\beq\label{DefFBPS}
d\Phi(p_a+p_b\rightarrow Q+p_r)=d\,\Phi(\widehat{p}_a+\widehat{p}_b\rightarrow Q)
\times d\,\Phi_{\mbox{\tiny FBPS}}(p_a,p_b,p_r)
\times\theta_{\mbox{\tiny veto}}\ ,
\eeq
where $\theta_{\mbox{\tiny veto}}$ (optionally) vetoes events that generate an additional jet.
At NLO the jet veto cut is simply,\begin{equation}
\theta_{\mbox{\tiny veto}}(p_r)=\theta\left[p^{{\tiny lab}}_T(p_r)<p_T^{\mbox{\tiny min}}(\mbox{jet})\right] \;,
\end{equation}
where $p^{{\tiny lab}}_T(p_r)$ is the laboratory frame
transverse momentum (calculated using Eq.~(\ref{eq:ptydef})).
Note the initial state brancher is necessarily an antenna brancher since it ensures that the initial state
partons remain massless.
The form of the FBPS generator, in terms of the kinematic variables $p_a$, $p_b$ and $p_r$, is,
\beq\label{eq:fbps}
d\,\Phi_{\mbox{\tiny FBPS}}(p_a,p_b,p_r)=
\frac{1}{(2\pi)^3}\left(\frac{\widehat s_{ab}}{s_{ab}}\right)d\,t_{ar}d\,t_{rb}d\,\phi\;,
\eeq
where
$t_{xy} = (p_x-p_y)^2$ and $d\phi$ is a rotational degree of freedom about the $z$-axis.
The explicit construction of the momenta $p_a$, $p_b$ and $p_r$ in terms of the integration variables is detailed 
in Appendix~\ref{app:FBPS}. The phase space weight corrects the flux factor due to the resulting emission
of an extra parton.

Finally, we observe that the forward brancher must by necessity change the initial state momenta. This means 
that for bremsstrahlung events the values of $p_T^{{\tiny lab}}$ will depend on the 
branching momentum $p_r$. Thus although 
the four momenta of the final state particles are fixed in the MEM frame the value of the $p_T^{{\tiny lab}}$ observable changes dynamically.
In other words a single event with fixed MEM frame four momenta corresponds to a range of  $p_T^{{\tiny lab}}$ values.
Using the FBPS we can now explicitly define $R_{\Omega}({\bf{x}})$ as,
\begin{eqnarray}
R_{\Omega}({\bf{x}}) = \int d\,\Phi_{\mbox{\tiny FBPS}}(p_a,p_b,p_r)\bigg( \mathcal{L}_{ij}(s_{ab},x_l,x_u)\mathcal{R}^{ij}_\Omega(p_a,p_b,{\bf{x}},p_r)\nonumber\\-\sum_{m}\mathcal{L}_{ij}(s_{ab},x^{m}_l,x^{m}_u)D^{m}(p_a,p_b,p_r)\mathcal{B}^{ij}_{\Omega}(\hat{p}_a,\hat{p}_b,{\bf{x}})\bigg).
\label{eq:Rdef}
\end{eqnarray}
In the above we note that the boost integral is defined for a given branching, since each branching generates a new $s_{ab}$.
The quantity $\mathcal{R}^{ij}_{\Omega}(p_a,p_b,{\bf{x}},p_r)=|M^{(0)}_{\Omega}(p_a,p_b,{\bf{x}},p_r)|^2$ is 
the Born level matrix element with one additional parton. Finally, $D(p_a,p_b,p_r)\mathcal{B}^{ij}_{\Omega}(\hat{p}_a,\hat{p}_b,{\bf{x}})$ represents the subtraction terms that cancel the soft and collinear 
divergences which occur when $p_r$ is unresolved. A couple of observations are in order in regards to the dipole pieces. We note that, since the dipoles
 must provide a pointwise cancellation, the boost function inherits the same $s_{ab}$ as in the real boost function. However the underlying Born matrix element 
 must be evaluated using the original Born $\hat s_{ab}$ in order to have a one-to-one correspondence with Eq.~(\ref{eq:kvirt}).
 This also fixes the integration limits, $x_l^m$ and $x_u^m$ in Eq.~(\ref{eq:Rdef}).  
 We discuss the exact modifications to the usual dipole subtraction scheme in Appendix~\ref{app:Subs}.

We are now in a position to build our scattering probability accurate to NLO, 
based on the quantities $V_{\Omega}({\bf x})$ and $R_{\Omega}({\bf x})$ that we have
defined in Eqs.~(\ref{eq:kvirt}) and~(\ref{eq:Rdef}) above. The NLO probability density function associated with the event ${\bf{x}}$ is,
\beq
\mathcal{P}({\bf{x}}|\Omega)=\frac{1}{\sigma^{NLO}_{\Omega}}\bigg(V_{\Omega}({\bf{x}})
+R_{\Omega}({\bf{x}}) \bigg)\;.
\label{eq:MEM_NLO}
\eeq 
This equation defines the MEM at NLO. 

\subsection{Generating unweighted events at NLO}  

A welcome by-product of the method outlined in the previous sub-section is its ability to 
generate unweighted events at NLO. In this section we outline how this is possible and in  
later sections we will use the technique to generate samples of unweighted events that can 
be used to test the MEM. 
The techniques described in this section is similar to the POWHEG method \cite{Frixione:2007vw}, which also 
projects NLO calculations onto Born phase spaces. However, the aim of this setup is not to provide a matched 
prediction, but a NLO event generator. Since our calculation is a NLO one, there is no guarantee that the event weights 
are positive. The method we will shortly describe is thus valid when the NLO calculation does not produce negative differential 
distributions. 

Our starting point is Eq.~(\ref{eq:Kfac}), in which we explicitly separated the NLO calculation into real and virtual contributions. 
We define the inclusive phase space spanned by the Born 
processes as $\Phi$, which we can separate into two regions. Region I is the part of the inclusive phase space, $\Phi$, that is 
populated by the LO calculation under the lab frame cuts. Region II is the remaining part of the inclusive phase space, in which 
the LO calculation does not contribute. 

We focus first on region I. Since the LO contribution is non-zero we can write a point by point $K$-factor as follows, 
\begin{eqnarray}
K_{I}({\bf x})&=&\frac{d\sigma^{NLO}}{d{\bf x}}\bigg(\frac{d\sigma^{LO}}{d{\bf x}}\bigg)^{-1} \nonumber \\
&=&\frac{V_{\Omega}({\bf x})+R_{\Omega}({\bf x})}{B_{\Omega}({\bf x})}.
\end{eqnarray}
This quantity is not positive definite since one can construct phase space points for which $K_{I}({\bf x}) < 0$. However, these correspond to 
regions in which the NLO calculation is unphysical. More specifically, it is possible to choose a renormalisation scale such that the differential 
cross section becomes negative. Typically this occurs because the choice of renomalisation scale is widely separated from the 
typical scale of the event. In general if a sensible scale choice is used then $K_{I}({\bf x}) > 0$. 
In order to ensure that $K_I({\bf x}) > 0$ it is sufficient to check that the NLO differential cross section is positive in all 
observables. 
One can then create weighted NLO events in this region by generating a Born phase 
space point and recording both the Born weight, $B_{\Omega}({\bf x})$ and the $K$-factor, $K_{I}({\bf x})$ for that point
(as well as the phase space weight associated with ${\bf x}$).  If the calculation is completely inclusive, i.e. no cuts are applied and region II is empty, 
then an unweighted NLO sample can easily be obtained by unweighting 
the combination of $K_{I}({\bf x})$, $B_{\Omega}({\bf x})$ and the phase space weight. 

In region II there is no $K$-factor since the LO cross section is zero.
In this region the virtual contribution, and all of the terms associated with the subtraction procedure, are zero since they occupy
the Born phase space. Hence $K_{II}({\bf x})$ is positive definite since it only corresponds to the LO process with an additional parton,
\begin{eqnarray}
K_{II}({\bf x}) = R_{\Omega}({\bf x}) \;.
\end{eqnarray}
Therefore in region II we construct our weights as a combination of the phase space weight associated with
${\bf x}$ and $K_{II}({\bf x})$. 

By combining regions I and II we have weights that span the entire phase space and which are positive
(with the caveat that the total NLO differential cross section should be positive everywhere).  
Although the events all have the structure of a Born phase space point, the sum over the associated weights results in the NLO cross section.
We stress that the events found in region II are those in which the Born contribution is zero due to fiducial cuts and 
not a kinematic cut off. For example if one demanded a leptonic $p_{T}$ cut of $15$~GeV then region II would correspond to $p_{T} < 15$ GeV.
On the other hand, if the lepton had some natural cut off (for example, $p_T > m_Z/2$) then this region is already excluded from the inclusive Born
phase space, $\Phi$.  Using the weighted sample described here one can produce 
unweighted events in exactly the same fashion as one does at LO. 

\subsection{Extension to MET} 

In order to have a MEM which works for all EW final states it is crucial to be able to include missing transverse energy (MET).
 Therefore in this subsection we introduce how MET can be included in the method. For simplicity we maintain 
our ``perfect detector'' setup. This is however, a crude approximation since the experimental resolution of MET is much worse 
than that of light charged leptons. However the aim of this section is to provide the {\it{theoretical}} definition of the MEM at NLO in the presence of MET. We 
leave the actual experimental analyses involving realistic transfer functions to future study. 

Introducing MET into the MEM clearly involves changing the factorization setup in eqs.~(\ref{eq:LO_XS})-(\ref{eq:boostMEM}), to reflect the 
lack of knowledge of the full final state.  Explicitly we now observe $m$ leptons plus MET where before we identified $m'$ leptons, as a result 
the definition of an observation becomes, 
\begin{eqnarray}
\prod_{i=1}^{m'} \delta^{(4)}(p_i-p^{obs}) \rightarrow \prod_{i=1}^{m} \delta^{(4)}(p_i-p^{obs})\delta^{(2)}(p_{T}-p^{obs}_{MET}).
\end{eqnarray}
Clearly the degrees of freedom are reduce by $m'-m+2$ relative to the fully identified final state. As a result our weight will entail additional integrations 
over these unobserved quantities, the number of which depends on the number of MET particles we insert into our Matrix Element model. 

For simplicity we present a detailed discussion of the case with one assumed neutrino and one lepton, i.e. $W$ production in the Standard Model. First we note 
that the total LO cross section can be written as, 
\begin{eqnarray}
\sigma^{LO}_{W\rightarrow \ell\nu} = (2\pi)^{-2} \int d x_a d x_b \delta(s x_a x_b - Q^2) \bigg(d^4 p_{\nu} \delta(p_{\nu}^2) \bigg) \frac{f_i(x_a)f_j(x_b)}{x_a x_b s} \mathcal{B}^{ij}_{W\rightarrow \ell\nu}
\end{eqnarray}
In obtaining this equation we have used the overall momentum conserving delta function to eliminate the integration over the electron phase space, the delta function in $x_a$ and $x_b$ arises
from the on-shell condition for the electron. The factorization we require for the MEM  involves contracting the above equation with the following transverse momentum constraining delta function, 
\begin{eqnarray}
\delta({\bf{x}}) = \delta(p^{obs}_{MET,x}-p^{\nu}_x) \delta(p^{obs}_{MET,y}-p^{\nu}_y) 
\end{eqnarray}
This freezes the transverse momentum, the on-shell constraint removes a further integration variable leaving the following integrand, 
\begin{eqnarray}
\sigma^{LO}_{W\rightarrow \ell\nu}\vert_{\delta({\bf{x}})} = (2\pi)^{-2} \int d x_a d x_b \delta(s x_a x_b - Q^2) \frac{d p^{\nu}_z}{E^{\nu}} \frac{f_i(x_a)f_j(x_b)}{x_a x_b s} \mathcal{B}^{ij}_{W\rightarrow \ell\nu}
\end{eqnarray}
This weight is nearly in the desired form, we use the delta function to eliminate the integration in $x_b$ and define our Born weight as 
\begin{eqnarray}
B^{LO}_{W\rightarrow \ell\nu}({\bf{x}}) = (2\pi)^{-2} \int d x_a  \frac{d p^{\nu}_z}{E^{\nu}} \frac{f_i(x_a)f_j(Q^2(p^{\nu}_z))}{Q^2(p^{\nu}_z) x_a} \mathcal{B}^{ij}_{W\rightarrow \ell\nu}
\end{eqnarray}
In the above we have made the explicit dependence of $Q^2$ (the invariant mass of the system) on $p_z$ clear\footnote{In practice it is sensible to make $Q^2$ the integration variable~\cite{Artoisenet:2010cn}.}. 
The differences with respect to the MEM with fully identified final states is clear, since one does not observe the full final state one must first generate the longitudinal degrees of freedom based on the model 
hypothesis, then for the generated invariant mass of the system one integrates over the equivalent longitudinal boosts in the same manner as for the identified final state.  This procedure naturally extends to 
NLO, i.e. one integrates over the longitudinal degrees of freedom for the neutrinos, and the FBPS. We provide an example of this type of process in the next section.

\section{Validation} 
\label{sec:valDY}

In this section we present a simple validation of the method outlined in the previous section,
focussing for simplicity
on the production of lepton pairs at the LHC, $pp\rightarrow Z/\gamma^* \rightarrow \ell^+\ell^-$.  
In the first instance we study physics in the MEM frame, comparing predictions for observables
in this frame with the more familiar ones obtained in the lab frame. For this exercise, we investigate
parton level calculations (at LO and NLO) and Pythia~\cite{Sjostrand:2007gs}. The use of Pythia is a valuable test
of our method since it contains the effects of a parton
shower, underlying event and hadronisation in its output. After this study we present a simple
comparison of the MEM method at LO and NLO, in the context of a $Z$ mass measurement.

\subsection{Physics in the MEM frame.} 

We begin by recalling the definitions of lab frame quantities ($p_T$ and $\eta$) that we use to apply cuts in the MEM frame,
\beqa
p_T^{{\tiny lab},i}&=&\sqrt{\frac{s_{ai}s_{ib}}{s_{ab}}} \;, \quad
\eta^{{\tiny lab},i}=\frac{1}{2}\log\left(\frac{x_a^2s}{s_{ab}}\frac{s_{ib}}{s_{ai}} \right) \;.
\eeqa 
In passing we note that, although it is not needed in the cases that we will discuss here, we can
also easily define lab-frame azimuthal angle and rapidity differences in a boost-independent fashion,
\begin{equation}
\Delta\eta_{ij}^{{\tiny lab}}=\frac{1}{2}\log\left(\frac{s_{bi} \, s_{aj}}{s_{ai} \, s_{bj}}\right) \;, \quad
\Delta\phi_{ij}^{{\tiny lab}}=\cos^{-1}\left(\cosh(\Delta\eta_{ij}^{{\tiny lab}})
 -\frac{s_{ij}}{2 p_T^{{\tiny lab},i} p_T^{{\tiny lab},j}}\right) \;.
\end{equation}
These definitions would be useful for more complicated processes that include jets or that
require the application of an isolation procedure.
We will also consider the MEM frame transverse momentum, which is defined in a more familiar way, 
\begin{eqnarray}
p^{MEM,i}_T=\sqrt{(p_i^x)^2+(p_i^y)^2} \;,
\end{eqnarray} 
where, of course, the four-vector $p^{\mu}$ is explicitly in the MEM frame. 
The MEM frame has no unique definition of pseudo-rapidity for a given event, since there are multiple frames connected by longitudinal boosts. 

\begin{figure}
\begin{center}
\includegraphics[scale=0.6,angle=0]{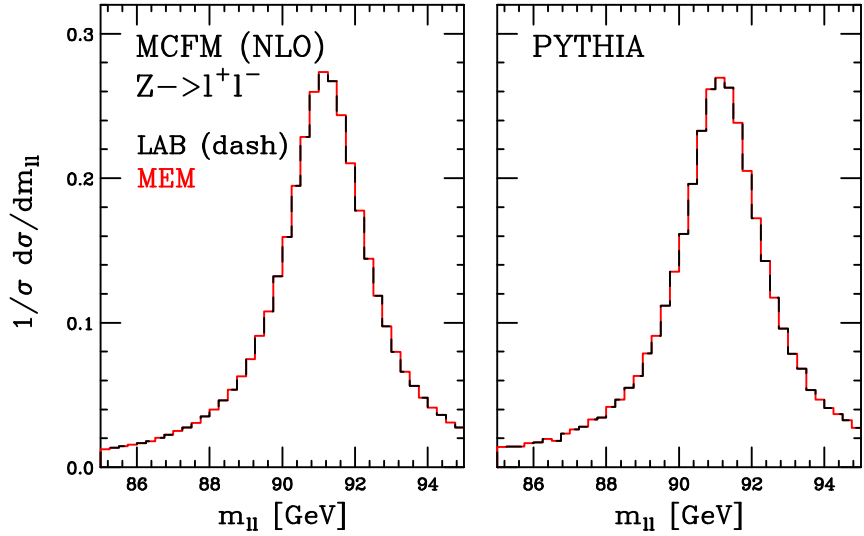} \vspace{0.2cm}\\ (a) \vspace*{0.5cm}\\ 
\includegraphics[scale=0.6,angle=0]{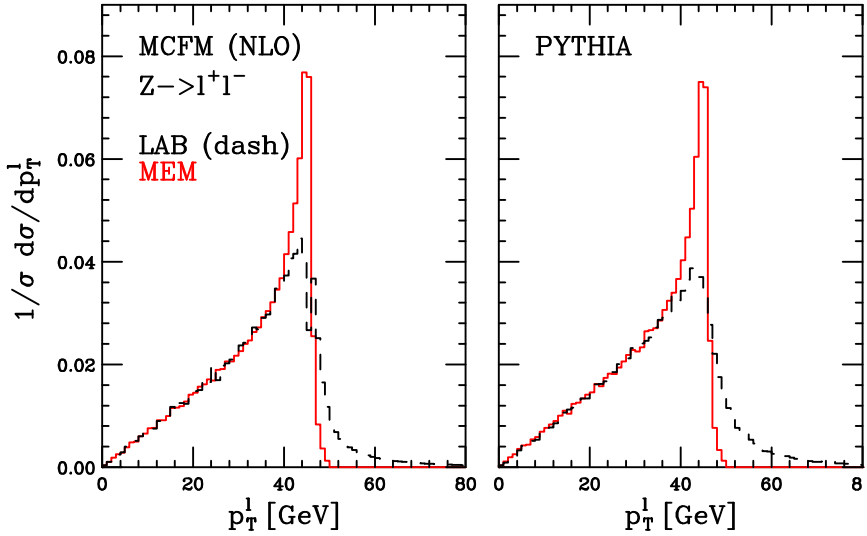} \vspace{0.2cm}\\ (b) \\ 
\caption{Comparison between the lab and MEM frame predictions from the NLO calculation of MCFM (left) and Pythia (right) for the process
$pp\rightarrow Z/\gamma^* \rightarrow \ell^+\ell^-$. In (a) we plot the invariant mass distribution of the two leptons and
in (b) we show the $p_T$ of the positively charged lepton. In each plot the lab frame quantity is shown in black (dashed), while the MEM frame
result is in red (solid).
\label{fig:memlab_comp}}
\end{center}
\end{figure}

We now wish to study the behaviour of different quantities in the lab and MEM frames.
We apply very loose cuts, namely we only require that the leptons lie in the invariant 
mass window, 
\begin{eqnarray} 
80~{\rm GeV} < m_{\ell^+\ell^-} < 100~{\rm GeV} \;.
\end{eqnarray} 
We generate LO and NLO parton level events using MCFM and more exclusive particle-level dilepton events using Pythia.
In Fig.~\ref{fig:memlab_comp} we compare the results from the lab and MEM frames for the quantities $p^{{\tiny lab}}_{T}$, 
$p^{MEM}_T$ and $m_{\ell\ell}$. 

In Fig.~\ref{fig:memlab_comp}(a) we see that, as is necessary, the invariant mass of the lepton pairs is identical in both frames. A more
interesting quantity is the frame-dependent $p_T$ of the positively charged lepton, $\ell^+$, shown in Fig.~\ref{fig:memlab_comp}(b).
At LO (parton level) the two quantities are the same because for pure LO results the
final state has zero net transverse momentum and thus the MEM and lab frames are identical. As soon as this simple picture is broken
the two frames are no longer the same and the $p_T$ distributions differ. This is apparent in both
the showered and NLO results. For the NLO and shower predictions it is possible, by radiating additional particles,
for a lepton to have lab-frame $p_T$ greater than $m_Z/2$. At LO this is not kinematically accessible, modulo small width effects.
This is demonstrated in the lab frame $p_T$ predictions for the NLO and showered results, shown in Fig.~\ref{fig:memlab_comp}(b), that
produce a high $p_T$ tail. The MEM frame, however,
requires that the event be boosted back to a Born topology. As such, the high $p_T$ region is not present  in this frame. Since the
overall normalisation  is fixed by the total cross section these events are manifested at lower values of $p_T$, with the region around $m_Z/2$
showing a considerable enhancement relative to the lab frame. 
We stress that the boost to the MEM frame has not changed the number or weight of each event, therefore the full NLO normalization 
is maintained.  

\begin{figure}
\begin{center}
\includegraphics[scale=0.75,angle=0]{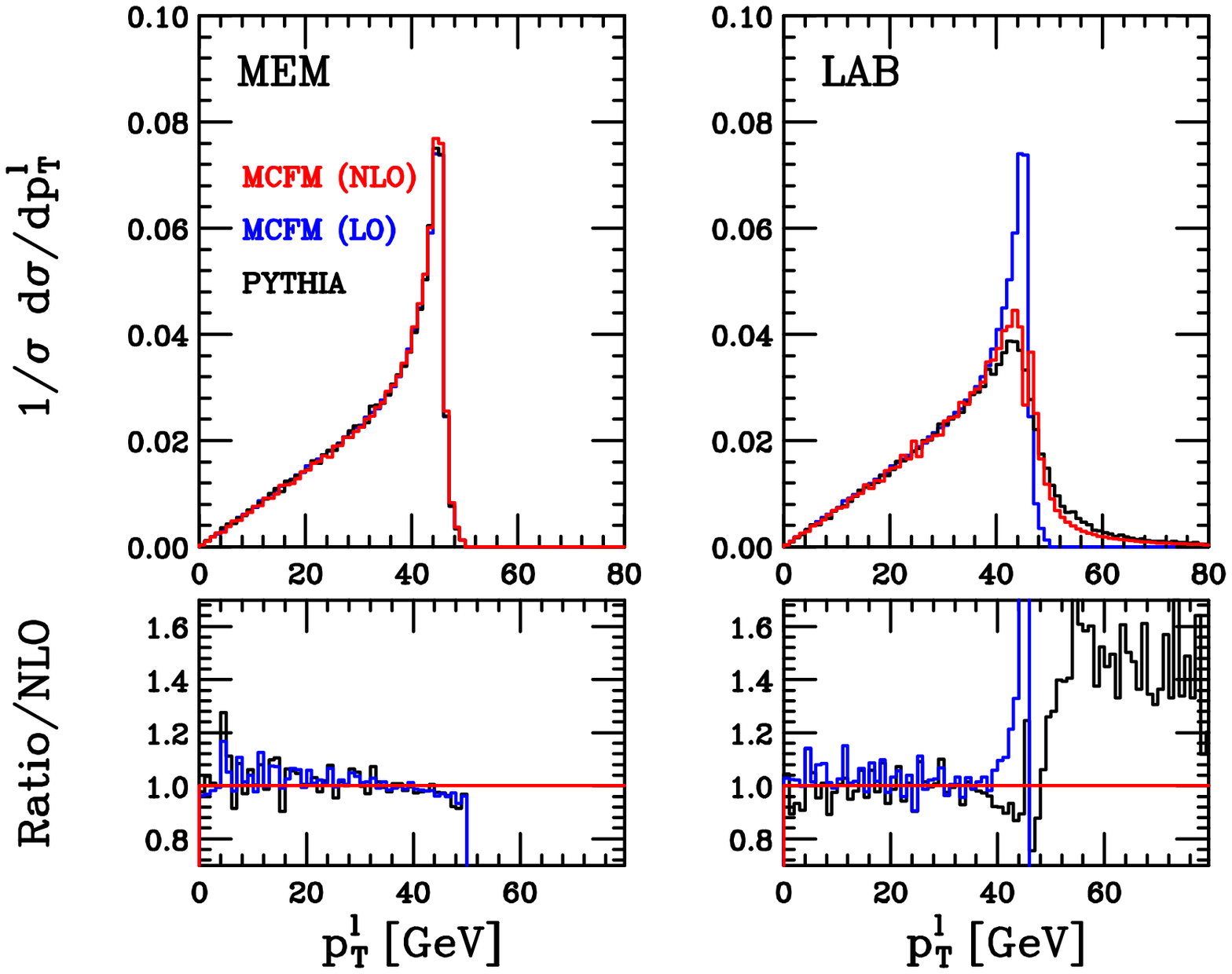} \vspace{0.2cm}
\caption{Comparison between MCFM (LO and NLO) and Pythia in different frames. On the left hand side $p_T^{\ell}$ is plotted in the MEM frame, whilst  on the right hand 
side the lab frame equivalent is plotted. Predictions are normalised by the total cross section (or number of events in the Pythia case). 
\label{fig:mcfmpy_comp}}
\end{center}
\end{figure}

In Fig.~\ref{fig:mcfmpy_comp} we directly compare the different theoretical predictions -- at LO, NLO
and using Pythia -- in both the lab and MEM frames.
It is clear  that the predictions in the MEM frame are very similar with respect to each other, with both LO and Pythia predicting a 
slightly softer spectrum relative to NLO. We note that the shape differences between NLO and the other predictions is
consistently of order 10\% or less. In the lab frame, however, there are significant differences between the predictions, in particular
in the region $p_T > m_Z/2$. From this  discussion we conclude that the MEM frame
possesses some very nice features. In particular the differences with respect to the LO prediction (from either shower or NLO) are consistent with 
naive estimates of higher order QCD effects,
suggesting good perturbative control. The main reason for the convergence is that, in the MEM frame,
kinematic ranges of observables are not extended beyond their LO boundaries.  
Since any such extension beyond the LO region is necessarily sensitive to further higher order corrections,
the elimination of this aspect of the calculation should be seen as an advantage for the MEM frame. 

\subsection{Validating the MEM: measuring $m_Z$}
In the previous sub-section we have used MCFM, representing a traditional approach to NLO calculations,
to generate lab frame events that are then transformed into the MEM frame. As described in the previous section, for the
the extension of the MEM method to NLO it is easiest to work directly in the MEM frame.
We have modified MCFM accordingly to incorporate the phase space generator and approach
described in the previous section.
In addition to the implementation of the FBPS,
the code has been constructed such that a NLO weight can be ascribed to an individual event in the MEM frame.

\begin{figure}
\begin{center} 
\includegraphics[scale=0.8,angle=0]{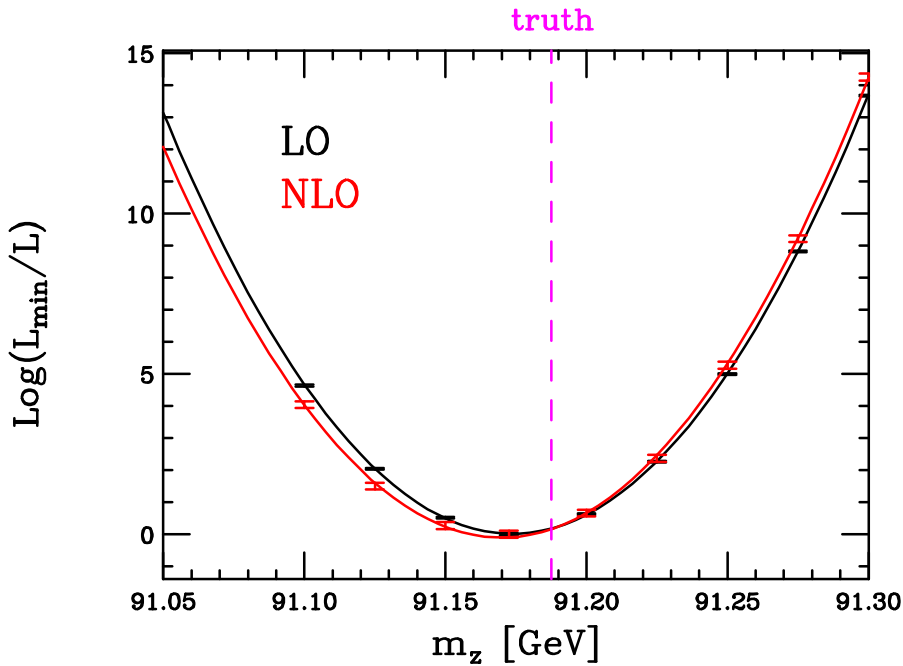} 
\caption{Log-likelihoods obtained by a MEM analysis at LO (black) and NLO (red) for the measurement of $m_Z$ at the LHC using Pythia data.
Errors represent MC integration uncertainty.} 
\label{fig:NLO_inc_Z}
\end{center}
\end{figure}

A simple test of our implementation of the MEM at LO and NLO is its application to the measurement of the mass of the $Z$ boson at the 7 TeV LHC. 
To this end we generate
${\mathcal{O}(5000)}$ events using Pythia that satisfy the following lab frame requirements, 
\begin{eqnarray}
p^{\ell}_{T} > 15~{\rm{GeV}} \;,  \quad  |\eta_{\ell}| < 2.5 \;, \quad
80~{\rm GeV} < m_{\ell^+\ell^-} < 100~\rm{GeV}. \;
\end{eqnarray} 
We use Pythia since it is a completely independent code to MCFM and as such is also independent of our new method for
generating the NLO weights. In addition, 
Pythia output is at the particle level, including shower, hadronisation and underlying event models. 
We note that in Pythia we have turned off both the mass of the leptons and QED radiation, both of which ensure our transfer function assumptions remain valid.  
In Fig.~\ref{fig:NLO_inc_Z} we present the likelihoods as a function of $m_Z$ for the completely inclusive case (i.e. the full data set).
As expected we observe a parabolic function around the best fit mass. Error bars represent the Monte Carlo integration uncertainty and statistical uncertainties can  be inferred by
using Eq.~(\ref{eq:nsig}). We observe that the truth value ($m_Z=91.1876$ GeV) easily lies within the 1-$\sigma$ band of our best fit values,
\beq
\mbox{LO:} \quad m_Z=91.170\pm 0.025~\mbox{GeV}\qquad \qquad  \mbox{NLO:} \quad m_Z=91.174\pm 0.025~\mbox{GeV}  \;.
\eeq
The power of the MEM  is also illustrated here, since with a data set of $\mathcal{O}(0.1)$ fb$^{-1}$ we are able to perform a measurement of the $Z$ mass
to within 25 MeV (modulo transfer function uncertainties). It is not surprising that the NLO and LO results are very close to one another
since we have already observed that, for this process, the NLO and LO kinematics are very similar in the MEM frame. 

\begin{figure}
\begin{center}
\includegraphics[scale=0.8,angle=0]{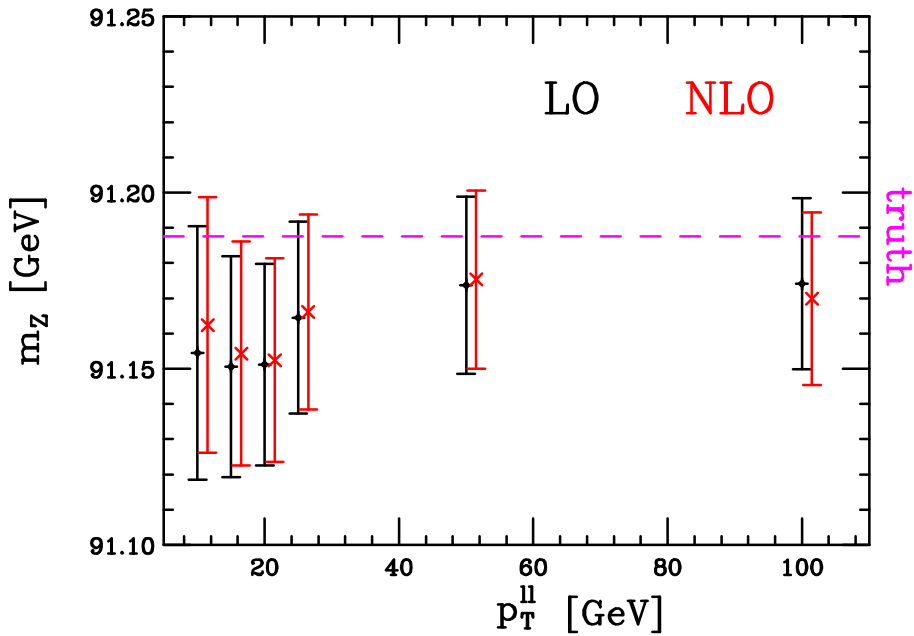} 
\caption{Reconstructed $Z$ mass as a function of the upper bound on the transverse momentum of the dilepton system, $p^{\ell\ell}_T$.
Errors represent the $ 1 \sigma$ deviation from the central value. Note that both LO and NLO calculations are performed at the
same values of the cut, $p^{\ell\ell}_T$. In the plot the NLO points have been moved slightly to the right for clarity.
\label{fig:ptll}}
\end{center}
\end{figure}

The results presented in Fig.~\ref{fig:NLO_inc_Z} are for the full sample that includes events in 
which there is a significant amount of showered radiation. Since there is no model of this additional radiation in the LO MEM, one may worry that the measured value
of $m_Z$ depends on the amount of this additional radiation. We therefore present the results of a study of this effect in Fig.~\ref{fig:ptll},
where we have performed the mass measurement for a variety of cuts on the transverse momentum of the dilepton ($Z$) system, $p_T^{\ell\ell}$.
By varying the maximum value of this quantity for events in our sample, we are limiting the amount of additional radiation
(i.e. showering) present in the event. Since this veto represents an  
additional cut on the data, the size of the data sample shrinks as the maximum $p_T^{\ell\ell}$ is reduced. For this 
reason the statistical uncertainty increases at low $p_T^{\ell\ell}$, as is apparent from the uncertainties shown in the figure. For this 
observable it is clear that both the dependence on the boost and on the higher order corrections is small. The relative independence of
the results from the amount of shower radiation allowed in the events 
illustrates that the boost method has worked well for this observable. This is encouraging 
but should not be taken as a general rule for all observables. The boost changes the parton fractions $x_a$ and $x_b$ and
thus observables that are sensitive to such changes will become dependent on the amount of additional radiation in the event.
In cases where imposing a jet veto is desirable, the boost (in)dependence should be checked by performing the measurement 
with a desired veto, and recalculating the observable with a tighter veto upon the same data set. If the two results agree within
statistical errors then one is  reassured that the shower is playing a minimal role. One may expect that, given its improved modeling of 
additional radiation, that the NLO results will be less sensitive to the additional radiation. 

\subsection{Example: measuring $m_W$} 

\begin{figure}
\begin{center} 
\includegraphics[scale=0.6,angle=0]{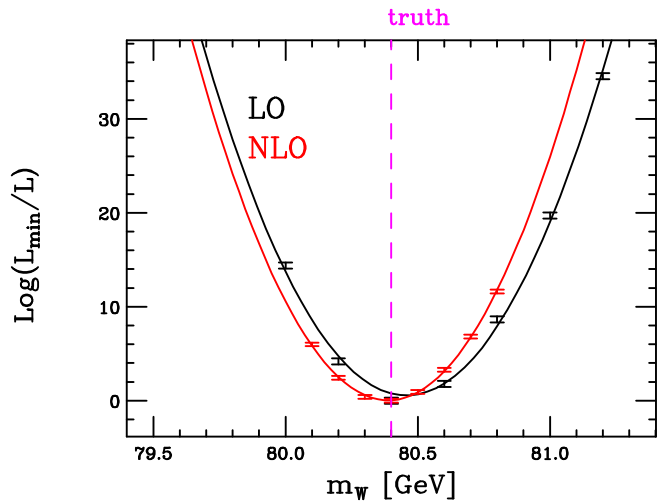} 
\caption{Log-likelihoods obtained by a MEM analysis at LO (black) and NLO (red) for the measurement of $m_W$ at the LHC using Pythia data.
Errors represent MC integration uncertainty.} 
\label{fig:massW}
\end{center}
\end{figure}

In order to illustrate the effects of including MET we present a simple example, namely measuring the $W$ mass. We generate 
4000 $W^+$ events using Pythia~\cite{Sjostrand:2007gs} for the LHC at 8 TeV. The final state must satisfy the following cuts, 
\begin{eqnarray}
p_T^{\ell^+} > 15  \,\,\rm{GeV} \quad |\eta^{\ell}| < 2.5  \quad    \rm{MET} > 15 \,\, {\rm{GeV}}. 
\end{eqnarray}
We do not include any kind of detector simulation, meaning that our neutrino transverse momentum is perfectly resolved. Whilst this is a crude 
experimental approximation it is a useful theoretical one, since we can directly compare our $W$ and $Z$ results and as such we will immediately see the effect 
of not observing the full final state and having to generate the longitudinal information as part of the model hypothesis. 
Our results are shown in Fig.~\ref{fig:massW}, corresponding to best fit values at  LO and NLO of, 
\beq
\mbox{LO:} \quad m_W=80.46\pm 0.09~\mbox{GeV}\qquad \qquad  \mbox{NLO:} \quad m_W=80.39\pm 0.08~\mbox{GeV}  \;.
\eeq
which should be compared to the truth value of $m_W=80.40$ GeV. It is interesting to note that the errors have increased by around a factor of 3  compared to the $Z$ measurements (for an event sample 
size is that is similar). One expects that in cases with more neutrinos the smearing becomes worse, since the constraints on the system (i.e. the measured missing energy) remain the same
whilst the number of degrees of freedom to be integrated over increases. 

\section{The Higgs Boson search in the channel $H \to ZZ\rightarrow 4\ell$ }\label{sec:ZZ}

A convenient example in which to test our MEM implementation at LO and NLO is the Higgs search at the LHC. 
One of the cleanest search channels is the process $H\rightarrow ZZ \rightarrow 4\ell$ \cite{ATLAS:2012ac,Chatrchyan:2012dg} since the final 
state can be fully reconstructed in the detector and the SM backgrounds are
small. With full control of the final state, with no sizeable missing transverse momentum or jet activity expected, this channel
is a natural candidate for a MEM approach.  The use of the MEM in this channel has been studied in some detail in ref.~\cite{Gainer:2011xz},
with the usual caveat of the leading order limitation. Since the NLO corrections to this process are large it is interesting to determine whether the MEM 
at NLO can improve upon the LO analysis.


In the following examples we will select events that contain four leptons satisfying the following
requirements, 
\begin{eqnarray} 
&&p_{T}^{\ell_1,\ell_2} > 20~{\rm{GeV}} \;, \quad
p_{T}^{\ell_3,\ell_4} > 5~{\rm{GeV}} \;, \quad
|\eta_{\ell}| < 2.0 \;, \nonumber\\
&&  15~{\rm GeV} < m_{\ell\bar\ell} < 115~{\rm GeV} \;, \quad 75~{\rm GeV} < m_{\ell'\bar\ell'} < 115~{\rm GeV}\;,
\label{eq:Higgscuts}
\end{eqnarray}
where leptons are labelled in order of decreasing transverse momentum from $\ell_1$ to $\ell_4$.
That is, we require one pair of oppositely-charged leptons to have an invariant mass within 
approximately $15$~GeV of the $Z$ mass while the invariant mass of the other pair is less constrained. 
In experimental searches the analysis cuts are typically tailored to the putative Higgs mass in
order to better discriminate against the relevant backgrounds. However, for simplicity, in our
studies we do not optimise the cuts in this way.
Therefore the limits and uncertainty ranges quoted here should be taken only as a rough estimate of
what can be achieved  in a true experimental analysis. Instead, we are more interested in assessing
the performance of the MEM at LO and NLO for a given set of cuts.

We perform our calculation for  the LHC operating at $\sqrt{s}=7$~TeV,
with $\mu_R=\mu_F=m_H$ in the calculation of the Higgs signal and $\mu_R=\mu_F=2m_Z$ for the $ZZ$ background.
We have used the CTEQ6 PDF set \cite{Lai:2010vv} matched to the appropriate order in perturbation theory.
Our NLO calculation includes the contributions from $gg\rightarrow ZZ$ for $n_f=5$ massless flavours, using results taken from MCFM~\cite{Campbell:2011bn}.
Although the interference between SM production of $WW$ pairs and the Higgs signal may be phenomenologically relevant~\cite{Campbell:2011cu}, in the $ZZ\rightarrow 4\ell$ channel the corresponding
interference is not expected to be important for a light Higgs boson since the final state is fully reconstructed. Although the interference effects may become non-negligible
for Higgs masses above a few hundred GeV, we do not include such effects here.

We begin by studying the scenario in which there is no Higgs boson and the only source of four lepton 
events is $pp\rightarrow ZZ\rightarrow 4\ell$ production, i.e. neglecting any other source of backgrounds.

We study the MEM using events samples which have been generated using SHERPA~\cite{Gleisberg:2008ta}, where a NLO calculation
has been matched to a parton shower and hadronization effects are also included. We then take the SHERPA input and boost it to the MEM frame as discussed in the previous sections. 
We note that in the MEM frame some of these events possess leptons with, for instance, $p^{MEM,\ell_4}_T < 5$~GeV. Since, at LO, $p^{MEM}_T=p^{lab}_T$ these events cannot pass the
fiducial cuts in the LO analysis and as such are not included in the calculation of the likelihood.
However, at NLO, the transverse momentum is not identical in the two frames, $ p^{MEM}_T \ne p^{lab}_T$. Therefore a value of $p^{MEM}_T < 5$~GeV can correspond to 
a real radiation contribution with $p^{lab}_T > 5$~GeV. As a result such events are included in the NLO likelihood calculation. Therefore there can be a different 
number of events in the LO and NLO data samples.  This is a reflection of the fact that the NLO calculation exhibits a richer kinematical structure than the LO one. 

In order that our assumption of an ideal detector is reasonable, we consider only Higgs bosons with masses of $300$~GeV or above. This
ensures that the width of the Higgs boson is sufficiently large (at least $8$~GeV) that the experimental detector resolution, embodied by the transfer functions,
should not be the dominant effect. To perform a realistic study in the region of lighter 
Higgs bosons would require detailed detector modeling of the transfer functions and is beyond the scope of this paper.

We generate pseudo-experiments based upon an expectation of 200 observed events. We then define our likelihood by,
\begin{eqnarray}
{\cal L}_{S+B}(\mu,N) = \frac{e^{-\mu} \mu^{N}}{N!}\prod_{i=1}^{N}\mathcal{P}({\bf x_i}| S=m_H) \;,
\end{eqnarray}
where $N$ is the number of events observed in the pseudo-experiment, and $\mu$ is the expected number of events for a given 
signal plus background hypothesis. This extended likelihood definition is more appropriate in the presence of signal and background 
contributions and when the number of events in each pseudo-experiment varies. In the presence of a signal hypothesis $\mathcal{P}$, the
weights that enter the likelihood are defined as,
\begin{eqnarray}
\mathcal{P}^{LO}({\bf x}_i| S=m_H) &=& \frac{1}{(\sigma^{LO}_{S}+\sigma^{LO}_{B})}\Bigl(B_{S}({\bf{x}}_i)+B_{B}({\bf{x}}_i)\Bigr) \;, \\
\mathcal{P}^{NLO}({\bf x}_i| S=m_H) &=&
 \frac{1}{\left(\sigma^{NLO}_{S}+\sigma^{NLO}_B\right)}\Bigl(V_{S}({\bf{x}}_i)+V_{B}({\bf{x}}_i)+ R_{S}({\bf{x}}_i)+R_{B}({\bf{x}}_i)\Bigr) \;.
\end{eqnarray} 
Note that we do not alter the expected number of events based upon the order in perturbation theory, i.e. we expect a 
background only hypothesis to generate 200 events in both our LO and NLO studies.
As a result, the LO hypothesis is not penalized by its lower prediction for the total rate relative to NLO. This procedure is
thus akin to rescaling the LO prediction for the rate to its NLO value.

\begin{figure}
\begin{center}
\includegraphics[scale=0.5]{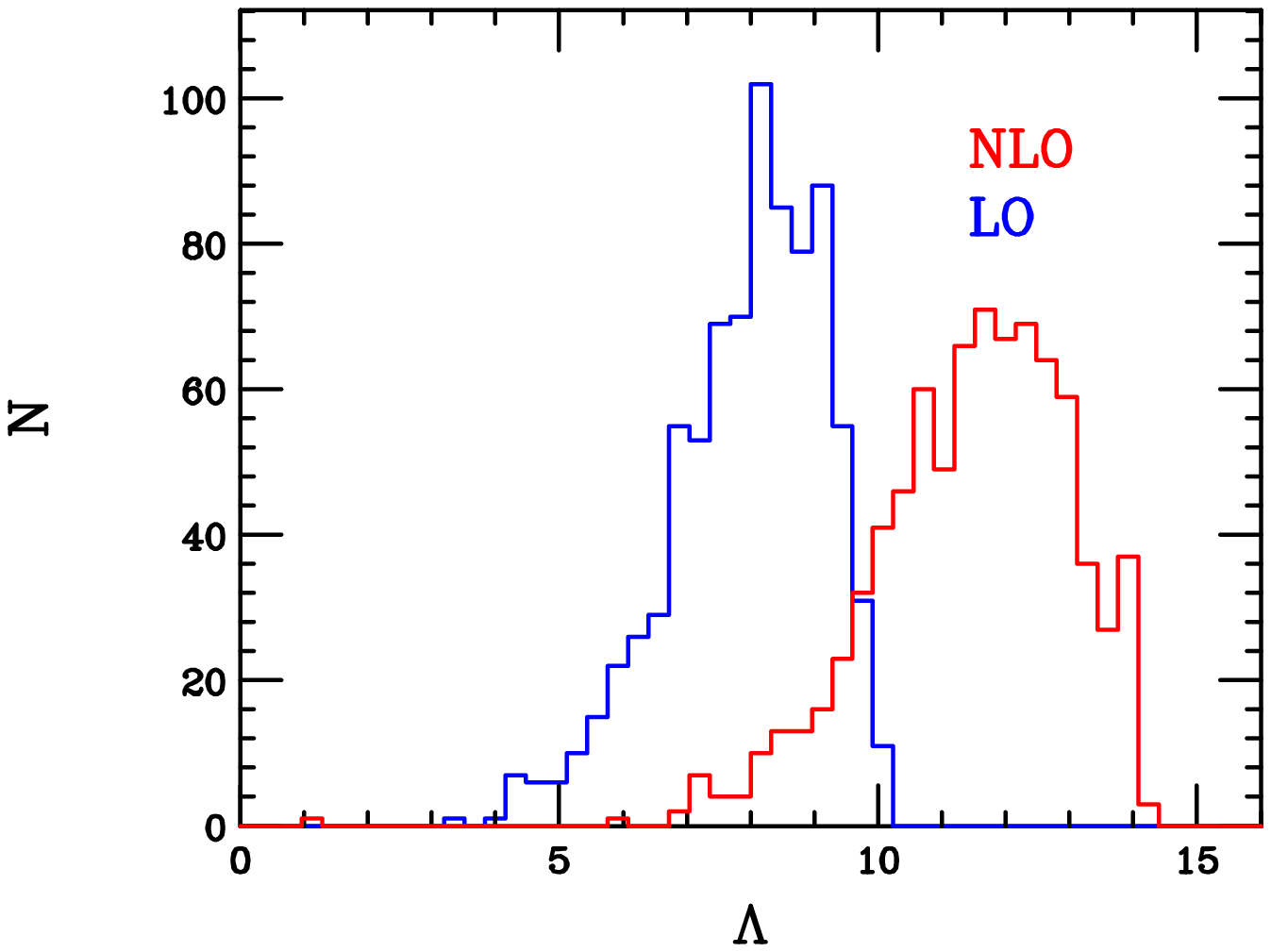}
\includegraphics[scale=0.5]{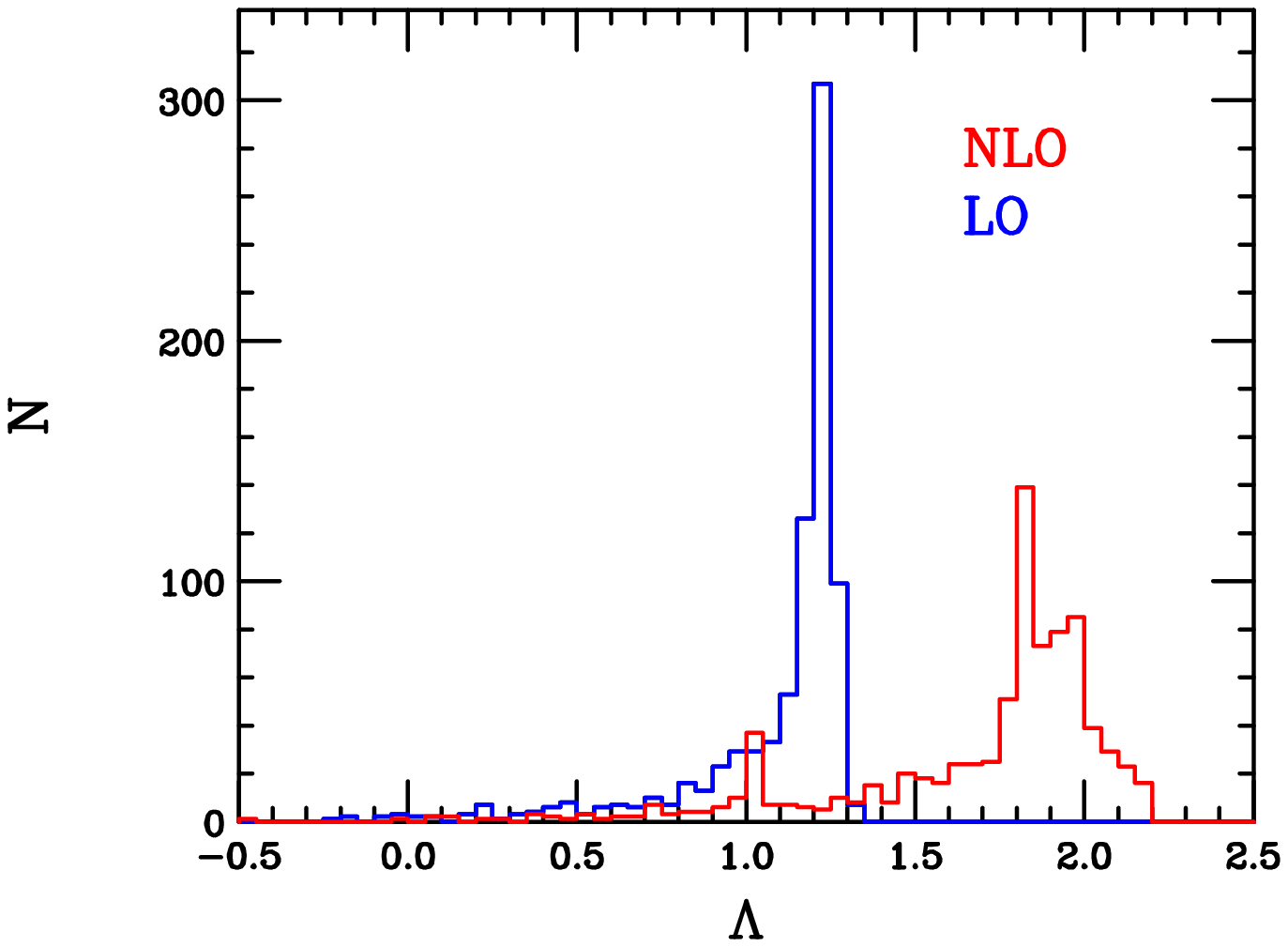}
\caption{Distribution of the log-likelihood difference, $\Lambda = \log({\cal L}_{B}/{\cal L}_{S+B})$ observed in 821 pseudo experiments
testing the hypothesis that there is a Higgs boson with $m_H=300$~GeV (left) and $m_H=550$~GeV (right).}
\label{fig:Higgshisto}
\end{center}
\end{figure}

We have performed 821 pseudo-experiments with the procedure outlined above, for Higgs mass hypotheses of $300$ and~$550$~GeV.
The results of these analyses are presented in Figure~\ref{fig:Higgshisto}, in terms of the 
log-likelihood difference, $\Lambda = \log({\cal L}_{B}/{\cal L}_{S+B})$.
Since the signal at $300$ GeV is relatively strong, a typical pseudo-experiment -- that contains only background events --
is able to exclude this hypothesis effectively, i.e. $\Lambda > 0$. We note that, as expected, the NLO MEM typically sets a much stronger exclusion than at LO
(the peak in the NLO distribution is in the region $\Lambda \sim 12$, whilst the LO peak is at $\Lambda \sim 8$).
From this ensemble we can calculate the expected value of $\Lambda$ for a typical pseudo-experiment, the
mean of the distributions in Figure~\ref{fig:Higgshisto}. Similarly, the standard deviation of the distribution gives
a measure of the spread of the expected results within the sample.

\begin{figure}
\begin{center}
\includegraphics[width=12cm]{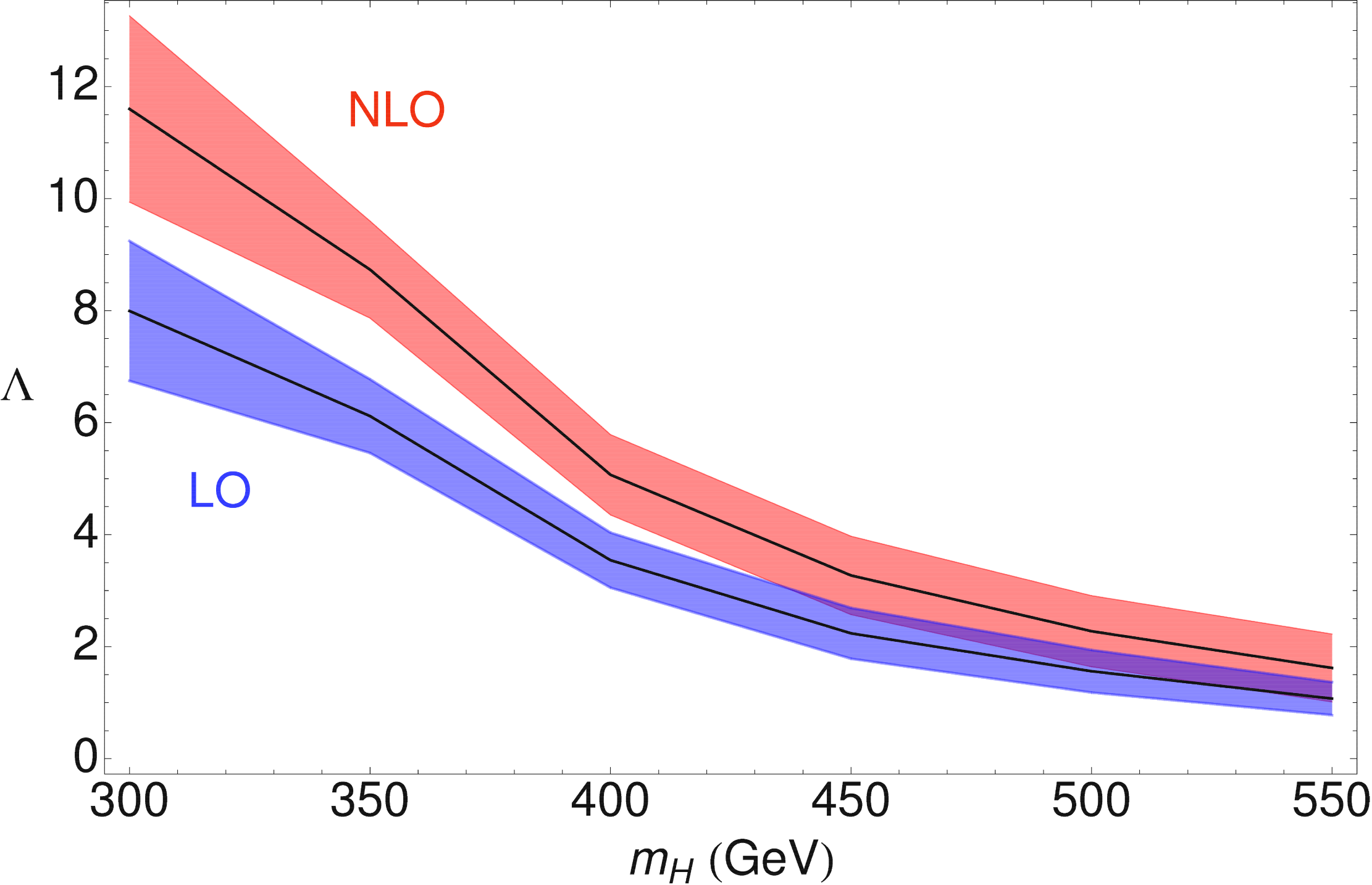}
\caption{The expected log-likelihood difference for background only and signal plus background, for a Higgs boson search in the channel, $H \to ZZ^\star \to 4$~leptons.
The red curves represent the NLO MEM results and the blue curve represents the LO MEM output. The shaded band indicates the 
standard deviation about the expectation.  }
\label{fig:HiggslogL}
\end{center}
\end{figure}

Repeating this exercise across the range $300$-$550$~GeV we obtain the results shown in 
Figure~\ref{fig:HiggslogL}, where we have indicated both the expected value of $\Lambda$ and the standard deviation of the
distribution. Note that the standard deviation, represented by the shaded band, should be 
treated only as a means of assessing the spread of results obtained by our method. It should not be interpreted as a rigorous definition of 
a confidence contour, such as one finds in an experimental analysis. 
We see that the pattern of results is repeated across the range of Higgs masses considered, with a significant difference
between the NLO and LO MEM results. The NLO method produces expected values of $\Lambda$ that
are consistently higher than at LO and are distributed with a larger standard deviation. However, the size of the standard
deviation relative to the expectated value of $\Lambda$ is similar at LO and NLO.
 As previously discussed, these differences cannot be attributed to a $K$ factor 
associated with an increased number of events arising at NLO.

\begin{figure}
\begin{center}
\includegraphics[width=12cm]{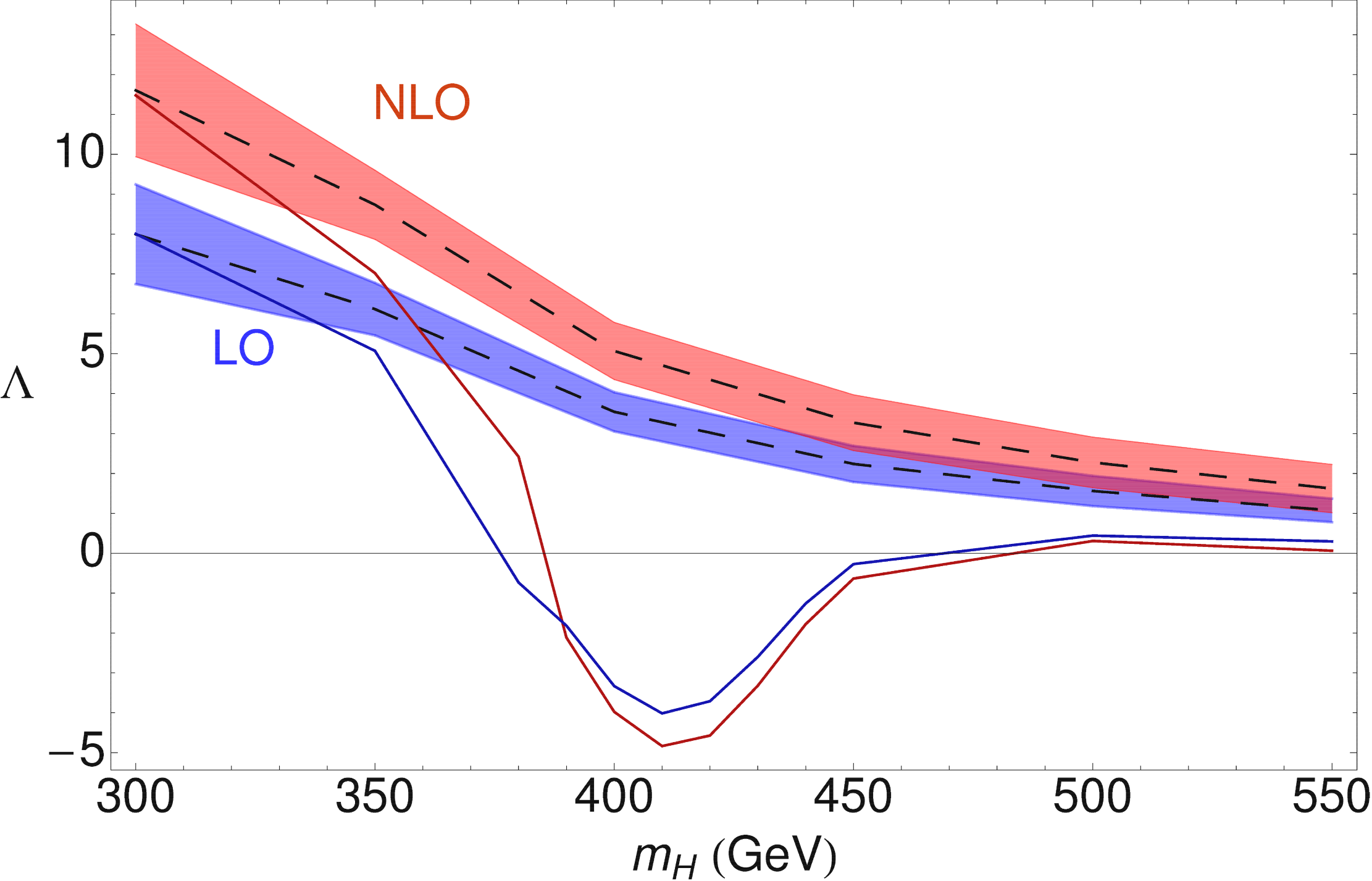}
\caption{Log likelihoods for a heavy Higgs boson decaying into four leptons, where we have injected a signal at $m_H=425$~GeV.
The expected log-likelihood, in the presence of no signal, is the same as in Figure~\protect\ref{fig:HiggslogL}.}
\label{fig:mh425}
\end{center}
\end{figure}

To investigate how the MEM performs in the presence of a genuine Higgs signal, we have added signal events to our sample
corresponding to a Higgs boson with mass $m_H=425$~GeV. We 
show our results with this signal injection in Fig.~\ref{fig:mh425}.
The deviation from the expected background-only result indicates that the sample with the injected signal cannot be
easily described by the background hypothesis. Moreover, the sample is
compatible with the Higgs signal hypothesis with $m_H$ in the $400$-$440$~GeV region, where $\Lambda<0$. 

\section{Conclusions}
\label{sec:conclusions}

The matrix element method is an analysis technique that can be used to determine parameters of an underlying physics model by using a set of
events that are measured experimentally. The probability that a single event in the set is described by a given
model hypothesis can be computed from a calculation of the scattering probability within that model. Up until now, the
use of this technique had been limited to scattering probabilities computed at the leading order in perturbation theory,
corresponding to Born matrix elements. 
In this paper we have illustrated how the method can be applied at NLO for electro-weak final states. 

Even at leading order, a key issue that must be addressed is the means by which a generic experimental
event is mapped to a scattering probability. In particular, such events typically contain additional hadronic 
activity that cannot be modelled by the simplest Born matrix elements. In this paper we have introduced
a procedure for handling this mapping in a consistent manner.
One can combine all of the event that is not part of the desired Born final state into
one four vector, $X$ and then boost into a frame in which $X$ is at rest in the transverse plane. This
feature does not uniquely define the boost and, although the matrix element is a Lorentz scalar,
the convolution with the parton distribution functions depends on the specific nature of the boost.
Therefore a theoretically well-defined procedure is only obtained by
integrating over all allowed boosts. Once this has been done we can produce a well defined LO weight
that can be associated with each experimental event. 

We have subsequently illustrated how one can extend this method to next-to-leading order. The incorporation
of some elements of the calculation, such as the virtual diagrams, is relatively straightforward since the
diagrams share the same phase space as the Born calculation. The inclusion of the real radiation contribution
is more complex and is performed by using a forward branching phase space generator. This allows one to maintain the
exact kinematics of the Born event whilst integrating out the real radiation.
We have used a slightly modified version of the usual Catani-Seymour dipole subtraction procedure
in order to ensure event-by-event subtraction scheme independence.
Using this generator we are able to define a map between all NLO events and Born phase space points.
The final result is a method
for generating a full NLO weight from a given Born phase space point. We note that there are some subtleties
in this method that require particular care. 
For example, the difference between the lab- and MEM-frame transverse momentum can mean that events that are within 
the fiducial region in the lab frame cannot ultimately be included in a LO MEM analysis. At NLO, such events can
be accommodated in a NLO MEM approach since they are accounted for by the presence of real radiation.

We have tested the method by producing NLO likelihoods for events that contain electroweak final states. As a first example
we considered production of lepton pairs and showed that one could correctly measure the mass of the $Z$ boson
using events generated with Pythia. In this instance we observed that the MEM frame kinematics are 
very similar at LO and NLO. For this reason we only observed small differences between the MEM at LO and NLO for this process. 

We then considered the search for the Higgs boson in the channel $gg \to H \to ZZ^\star \to 4$~leptons.
We showed that in this case there were differences between the LO and NLO MEM analyses, when analyzing a sample
of pseudo-data generated with the SHERPA code.
Statements regarding possible improvements in results
from using the MEM at NLO are difficult to make without further studies using transfer functions
based on a more realistic experimental setup. At the very least, using the MEM at LO and NLO gives a greater
control of systematic uncertainties arising from the perturbative expansion.

Future applications of the NLO MEM are widespread. One obvious example is the measurement of the top quark
mass. In addition the MEM  is very useful when data samples are limited by statistics. Examples of measurements
that would fall into this category include the measurement  of the properties of the Higgs boson and limits on
anomalous gauge boson couplings. We hope to extend our method to include more complicated final states, such as 
ones containing neutrinos and jets, shortly. The examples presented here have been implemented in a Fortran
code that may be obtained from the authors on request.\footnote{
Please contact {\tt ciaran@fnal.gov} for a copy of the code.}

\section*{Acknowledgments} 

We thank Oleg Brandt, Keith Ellis, Konstantin Matchev, Olivier Mattelaer and Gerben Stavenga for useful discussions.
We thank Adam Martin for help with the Pythia sample generation. 
Fermilab is operated by Fermi Research Alliance, LLC under Contract No.
DE-AC02-07CH11359 with the United States Department of Energy.

\appendix

\section{An Initial State Forward Branching Phase Space Generator} \label{app:FBPS}

In this appendix we will discuss the generation of the forward branching phase space (FBPS) used 
in our method, additional details for which can be found in refs.~\cite{Giele:1993dj,Giele:2011tm}.
To start the derivation of the initial state FBPS generator we 
recall the phase space for the production of a heavy state $Q$ 
from two initial partons, $\widehat p_a$ and $\widehat p_b$, as,
\beq\label{bornPS}
\frac{1}{2\widehat s_{ab}} d\,\Phi^{[D]}_1(\widehat{p}_a+\widehat{p}_b\rightarrow Q)=
\frac{2\pi}{2\widehat{s}_{ab}}\delta(\widehat{s}_{ab}-Q^2)\ .
\eeq
Here we have maintained the notation used in sec.~\ref{sec:MemNLO}, where hatted momenta 
indicate the underlying Born topology and unhatted momenta represent the phase space with 
one additional parton. 
The $D$-dimensional phase space for the emission of one extra initial state parton with momentum $p_r$
is given by~\cite{Giele:1993dj},
\beqa
\frac{1}{2s_{ab}}d\,\Phi^{[D]}_1(p_a+p_b\rightarrow Q+p_r) 
&=&\frac{(2\pi)^{1-D}}{4}
\left(\frac{\widehat s_{ab}}{s_{ab}^2}\right)
\left(\frac{t_{ar}t_{rb}}{s_{ab}}\right)^{(D-4)/2}
\!\!\!\!d\,t_{ar}d\,t_{rb}d\Omega^{[D-3]} \nonumber\\
&&\times\left[\frac{2\pi}{2\widehat s_{ab}}d\,\Phi^{[D]}_1(\widehat{p}_a+\widehat{p}_b\rightarrow Q)\right]\nn
&=&\frac{1}{2s_{ab}}d\,\Phi^{[D]}_{\mbox{\tiny FBPS}}\times
\left[\frac{2\pi}{2\widehat s_{ab}}d\,\Phi^{[D]}_1(\widehat{p}_a+\widehat{p}_b\rightarrow Q)\right]\ .
\eeqa
We need the FBPS generator in four-dimensions,
\beq\label{FBPS}
d\,\Phi^{[4]}_{\mbox{\tiny FBPS}}=d\,\Phi_{\mbox{\tiny FBPS}}=
\frac{1}{(2\pi)^3}\left(\frac{\widehat s_{ab}}{s_{ab}}\right)d\,t_{ar}d\,t_{rb}d\,\phi\ ,
\eeq
where $\phi$ is the azimuthal angle around the $z$-axis, $s_{xy}$ and $t_{xy}$ are invariants defined through, 
$s_{xy}=(p_x+p_y)^2$ and $t_{xy}=(p_x-p_y)^2$.
While the above formula gives the phase space integrator, we need to derive both the integration boundaries
and the explicit construction of the generated four-vectors ($p_a$, $p_b$ and $p_r$) that
are used in a numerical Monte Carlo integrator.
The phase space generator starts using the input momenta $\widehat p_a$ and $\widehat p_b$,
\beq
\widehat p_a=\widehat E_a\left(1,0,0,-1\right) \;, \qquad 
\widehat p_b=\widehat E_b\left(1,0,0,1\right) \;,
\eeq
with $\widehat s_{ab}=2\,\widehat p_a\cdot\widehat p_b=4\,\widehat E_a\widehat E_b$.
We can eliminate $t_{ar}$ in favour of $s_{ab}$ by inserting the operator
$\int d\,s_{ab}\ \delta(s_{ab}+t_{ar}+t_{rb}-\widehat s_{ab}) = 1$
to perform the $t_{ar}$ integration, 
\beq
d\,\Phi_{\mbox{\tiny FBPS}}=\left(\frac{\widehat s_{ab}}{(2\pi)^3}\right)
\int_{-t_{\mbox{\tiny min}}}^0d\,
t_{rb}\int_{\widehat s_{ab}}^{s}
\left(\frac{d\,s_{ab}}{s_{ab}}\right)
\int_0^{2\pi}d\,\phi\ .
\eeq
The integration limits on $s_{ab}$ can be understood from the momentum conserving delta function and 
the requirement that $t_{ar}$, $t_{rb} < 0$. We will define $t_{\mbox{\tiny min}}$ shortly. 
Our task is then to construct the new momenta $p_a$, $p_b$ and $p_r$ 
from the MC integration variables and
determine the integration boundary $t_{\mbox{\tiny min}}$. 
We relate $s_{ab}$ and $t_{rb}$ to our MC integration variable using logarithmic sampling,
\beqa
\int_{\widehat{s}_{ab}}^{s}\frac{d\,s_{ab}}{s_{ab}}
&=&\log\left(\frac{s}{\widehat{s}_{ab}}\right)\int_0^1 d\,r;\
s_{ab}(r)={\hat{s}_{ab}}^r s^{1-r} \\
\int_{-t_{\mbox{\tiny min}}}^0d\,t_{br}&=&
\int_{-t_{\mbox{\tiny min}}}^{-t_{\mbox{\tiny soft}}}d\,t_{br}
+\int_{-t_{\mbox{\tiny soft}}}^0d\,t_{br} \nonumber\\
\int_{-t_{\mbox{\tiny min}}}^{-t_{\mbox{\tiny soft}}}d\,t_{br}&=&
\log\left(\frac{t_{\mbox{\tiny min}}}{t_{\mbox{\tiny soft}}}\right)\int_0^1d\,r\ t_{rb}(r);\ 
t_{rb}(r)=-\left(t_{\mbox{\tiny min}}\right)^r\left(t_{\mbox{\tiny soft}}\right)^{1-r}\ .
\eeqa
Our phase space measure is now written in terms of MC integration variables and
our final task is to determine $p_a$, $p_b$ and $p_r$ for use in the matrix element 
in terms of our new variables. 
We wish to branch one of our initial state momenta and in this example we choose to branch 
$\widehat{p}_b$.
In order to do so we have to give it a virtuality $t_{rb}$, which we can do
by boosting $\widehat p_a$,
\beq
\tilde p_a=(1+\beta)\,\widehat p_a \;, \qquad
\tilde p_b=\widehat p_b-\beta\,\widehat p_a \;,
\eeq
with $\beta=-t_{rb}/\widehat s_{ab}$. Note that
$\widehat p_a+\widehat p_b=\tilde p_a+\tilde p_b=p_a-p_r+p_b$.
This means we have added to the phase space generator
a factor $\int d\,\beta\ \delta(\beta+t_{rb}/\widehat{s}_{ab})$ that does not change
the phase space weight.
We define $p_a=\tilde p_a=(1+\beta)\,\widehat p_a$ and
parametrize $p_b$ as follows,
\beq
p_b=z\,\widehat E_b\left(1,\cos\theta,\sin\theta\cos\phi,\sin\theta\sin\phi\right)\ ,
\eeq
where $\theta$ is the polar angle with respect to momentum $\widehat p_b$.
Momentum conservation now fixes $p_r=p_b-\tilde p_b=p_b-\widehat p_b+\beta\widehat p_a$.
To express $\cos\theta$ and $z$ in terms of the integration variables
and the input energies we calculate the invariants
\beqa
&&\left\{
\begin{array}{l}
t_{ar}=(p_a-p_r)^2=(\widehat p_a+\widehat p_b-p_b)^2\\
t_{rb}=-2\,p_r\cdot p_b=2\,\widehat p_b\cdot p_b-2\,\beta\widehat p_a\cdot p_b
=2\,\widehat p_b\cdot p_b+2\,(t_{rb}/\widehat s_{ab})\,\widehat p_a\cdot p_b
\end{array}
\right.\nn
&\Rightarrow&\left\{
\begin{array}{l}
t_{ar}=2\,\left(\widehat p_a\cdot\widehat p_b-\widehat p_a\cdot p_b-\widehat p_b\cdot p_b\right)\\
t_{rb}=2\,(\widehat p_a\cdot\widehat p_b)\,(\widehat p_b\cdot p_b)/(\widehat p_a\cdot\widehat p_b-\widehat p_a\cdot p_b)
\end{array}
\right.\nn
&\Rightarrow&\left\{
\begin{array}{l}
t_{ar}=4\,\widehat E_a\widehat E_b-2\,z\widehat E_b(\widehat E_a+\widehat E_b+(\widehat E_a-\widehat E_b)\cos\theta)\\
t_{rb}=4\,z\widehat E_b^2(1-\cos\theta)/(2-z(1+\cos\theta))
\end{array}
\right.\ .
\eeqa
We can invert the equations to obtain.
\beqa
z&=&\frac{4\,\widehat E_b^2(\widehat s_{ab}-t_{ar}-t_{rb})+t_{ar}t_{rb}}{4\,\widehat E_b^2(\widehat s_{ab}-t_{rb})}
  =\frac{4\,\widehat E_b^2s_{ab}+t_{ar}t_{rb}}{4\,\widehat E_b^2(\widehat s_{ab}-t_{rb})}\nn
\cos\theta&=&\frac{4\,\widehat E_b^2(\widehat s_{ab}-t_{ar}-t_{rb})-t_{ar}t_{rb}}
{4\,\widehat E_b^2(\widehat s_{ab}-t_{ar}-t_{rb})+t_{ar}t_{rb}}
=\frac{4\,\widehat E_b^2s_{ab}-t_{ar}t_{rb}}{4\,\widehat E_b^2s_{ab}+t_{ar}t_{rb}}\ .
\eeqa
By choosing 
\[
t_{\mbox{\tiny min}}=\min(s_{ab}-\widehat s_{ab},\widehat s_{ab}(\sqrt{s}-\widehat E_a)/\widehat E_a)\ ,
\]
we fulfill the requirement $-1<\cos\theta<1$. Remaining constraints (such as a jet-veto on $p_r$) are imposed through event vetos. 

We have thus illustrated how we have implemented the FBPS to perform both the integration over 
emitted partons and the phase space generation of $p_a$, $p_b$ and $p_r$ for use in the matrix element. 
The input for the generator is just the Born kinematics, i.e. $\widehat{p}_a$, $\widehat{p}_b$ and $Q$.

\section{Subtraction terms in the MEM frame} 
\label{app:Subs} 

In this appendix we discuss the modifications to the Catani-Seymour dipoles~\cite{Catani:1996vz} needed to correctly ensure
a one-to-one map between the integrated and unintegrated subtractions on an event by event basis.  In this paper we consider
processes with electroweak final states, and as such only need initial-initial dipoles. In the standard approach one would
perform a transformation such that the emitter and spectator are kept along the beam axis, with  a Lorentz transformation on
the remaining final state particles performed in order to ensure momentum conservation. In our case it is essential to
keep the final state particles fixed and instead change the momenta of the initial state partons. 

The standard Catani Seymour dipole keeps the momentum of the spectator initial particle $b$ fixed, while 
the emitter $a$ is rescaled by an amount $x_{a,r}$, 
\begin{eqnarray}
\widetilde{p}_{ar}&=&x_{a,r} \, p_{a} \;, \nonumber \\
x_{r,ab}&=&\frac{s_{ab}+s_{ar}+s_{rb}}{s_{ab}} \;.
\label{eq:CSdipole}
\end{eqnarray} 
Here we have kept the same notation as the previous section, with $r$, $a$ and $b$  representing the emitted parton, 
initial state emitter and initial state spectator respectively. Hatted momenta still represent 
the underlying Born phase space -- with unhatted momenta indicating the real phase space point -- and in addition
$\tilde{p}$ now represents the dipole phase space point. The transformation above is given by Eqs.~(5.137) and (5.138) in 
Ref.~\cite{Catani:1996vz} using our momentum definitions.  In order to ensure that $\widetilde{p}$ is a correct phase space 
point one must perform a Lorentz transformation (Eqs.~(5.139) - (5.144) in Ref.~\cite{Catani:1996vz})
to ensure momentum conservation. 
  
The above transformation is not ideal for our setup. This is because the Lorentz transformation 
will naturally change the underlying Born phase space point. This means that there will not 
be a one-to-one correspondence between real and virtual events and only the sum over all virtual and real contributions
will be well-defined. In order to maintain our exact map to the Born phase space $\hat{p}_a+\hat{p}_b\rightarrow Q$ we 
replace Eq.~(\ref{eq:CSdipole}) by the following  transformation, 
\begin{eqnarray}
\widetilde{p}_{ar}&=&x_{a,r} \, \hat{p}_{a} \;, \\
x_{r,ab}&=&\frac{s_{ab}+s_{ar}+s_{rb}}{s_{ab}} \;.
\end{eqnarray}
Note that the transformation acts on $\hat{p}_{a}$, the initial state momentum of the Born phase space. We note 
that this transformation preserves momentum conservation in the transverse plane, but 
not in the longitudinal plane. Therefore the correct dipole phase space point is at a different 
$x_a$ and $x_b$ than the original Born phase space point. Since we integrate over these variables 
this is sufficient to obtain the exact mapping between virtual and real contributions on am event by event basis.  
Using our new transformation we can implement the usual Catani-Seymour dipole formulae
(Eqs.~(5.145) - (5.156) in Ref.~\cite{Catani:1996vz}).

\bibliography{CGW}
\bibliographystyle{JHEP}

\end{document}